\documentclass[twocolumn,preprintnumbers,prd,superscriptaddress,nofootinbib,floatfix,showpacs,showkeys]{revtex4-2}

\usepackage{amsmath}
\usepackage{amsfonts}
\usepackage{amssymb}
\usepackage{graphicx}
\usepackage{color}
\usepackage{xcolor}
\usepackage{hyperref}
\usepackage{booktabs}
\usepackage{hyperref}
\usepackage{cleveref}
\usepackage{braket}
\usepackage{mathtools}
\usepackage[rightcaption]{sidecap}
\usepackage{soul}
\usepackage{enumitem}
\usepackage{comment}
\usepackage{physics}
\usepackage{subfigure}
\usepackage{dsfont}

\definecolor{vividviolet}{rgb}{0.62, 0.0, 1.0}
\definecolor{amaranth}{rgb}{0.9, 0.17, 0.31}
\definecolor{palatinateblue}{rgb}{0.15, 0.23, 0.89}
\definecolor{brightpink}{rgb}{1.0, 0.0, 0.5}
\definecolor{cornflowerblue}{rgb}{0.39, 0.58, 0.93}
\definecolor{deepcarminepink}{rgb}{0.94, 0.19, 0.22}
\definecolor{radicalred}{rgb}{1.0, 0.21, 0.37}
\hypersetup{ linktoc=all,
    colorlinks, linkcolor={palatinateblue},
    citecolor={brightpink}, urlcolor={amaranth}
}

\newcommand{\be}{\begin{equation}}
\newcommand{\ee}{\end{equation}}
\newcommand{\bs}{\begin{split}}
\newcommand{\bea}{\begin{eqnarray}}
\newcommand{\eea}{\end{eqnarray}}

\newcommand{\bes}{\begin{subequations}}
\newcommand{\ees}{\end{subequations}}

\newcommand{\xx}{\mathbf{x}}

\newcommand{\kk}{\boldsymbol{k}}
\newcommand{\x}{\mathsf{x}}

\usepackage{tikz,xcolor}
\definecolor{lime}{HTML}{A6CE39}
\DeclareRobustCommand{\orcidicon}{%
	\begin{tikzpicture}
	\draw[lime, fill=lime] (0,0) 
	circle [radius=0.16] 
	node[white] {{\fontfamily{qag}\selectfont \tiny ID}};
	\draw[white, fill=white] (-0.0625,0.095) 
	circle [radius=0.007];
	\end{tikzpicture}
	\hspace{-2mm}
}
\foreach \x in {A, ..., Z}{%
	\expandafter\xdef\csname orcid\x\endcsname{\noexpand\href{https://orcid.org/\csname orcidauthor\x\endcsname}{\noexpand\orcidicon}}
}

\begin{document}
\title{Relativistic quantum communication between harmonic oscillator detectors}

\author{Alessio Lapponi\orcidA{}}
\email{alessio.lapponi-ssm@unina.it}
\affiliation{Scuola Superiore Meridionale (SSM), Largo San Marcellino, Napoli, 80138 Italy.}
\affiliation{Istituto Nazionale di Fisica Nucleare (INFN), Sezione di Napoli, Napoli, 80126, Italy.}

\author{Dimitris Moustos\orcidB{}}
\email{dmoustos@upatras.gr}
\affiliation{Department of Physics, University of Patras, 26504 Patras, Greece}

\author{David Edward Bruschi\orcidC{}}
\email{david.edward.bruschi@posteo.net}
\affiliation{Institute for Quantum Computing Analytics (PGI-12), Forschungszentrum J\"ulich, 52425 J\"ulich, Germany}

\author{Stefano Mancini\orcidD{}}
\email{stefano.mancini@unicam.it}
\affiliation{School of Science and Technology, University of Camerino, Via Madonna delle Carceri, Camerino, 62032, Italy.}
\affiliation{Istituto Nazionale di Fisica Nucleare (INFN), Sezione di Perugia, Perugia, 06123, Italy.}

\begin{abstract}
We propose a model of communication employing two harmonic oscillator detectors interacting through a scalar field in a background Minkowski spacetime. In this way, the scalar field plays the role of a quantum channel, namely a Bosonic Gaussian channel. The classical and quantum capacities of the communication channel are found, assuming that the detectors' spatial dimensions are negligible compared to their distance. In particular, we study the evolution in time of the classical capacity after the detectors-field interaction is switched on for various detectors' frequencies and coupling strengths with the field. As a result, we find a finite value of these parameters optimizing the communication of classical messages. Instead, a reliable communication of quantum messages turns out to be always inhibited.
\end{abstract}

\vspace{0.5cm}


\maketitle

\section{Introduction}
Quantum communication is one of the preeminent applications of quantum information theory \cite{Gisin,wilde2017quantum,mancini2019quantum}. Quantum communication, in the broader sense, is concerned with the transfer of quantum states  through a quantum channel. Such states are usually employed to encode quantum information that must be shared between two or more users. With the rapid development of space-based quantum technologies \cite{kaltenbaek2021quantum,sidhu2021advances,belenchia2022quantum}, which require the exchange of photons between distant users via satellite nodes \cite{liao2017satellite,yin2017satellite,Pan2018,Dequal:Villoresi,Vallone:Villoresi,Calderaro_2018}, reliable transmission of quantum states over long distances  becomes important. Since operations in space are inherently affected by motion \cite{Bruschi:cavity:1,Bruschi:cavity:2,Bruschi:cavity:3,Bruschi:cavity:4,Bruschi:cavity:5}  and gravity \cite{Bruschi:Satellite,Kohlrus:Bruschi,Exirifard:Karimi:1,Exirifard:Karimi:2}, it is of current interest to understand how  relativistic motion of physical system or the curvature of the background spacetime affect the transmission of quantum information.

Relativistic quantum communication channels extend their purely quantum counterparts to regimes where relativity plays a role \cite{Cliche:Kempf2010}. For example, non-static spacetimes can be considered as relativistic quantum channels since the information transmission is affected by the spacetime evolution \cite{Mancini_2014,Good_2021}. In the context of static spacetimes, such channels can be used to study the transmission of information between ideal pointlike two-level quantum probes known as Unruh-DeWitt (UDW) particle detectors that is mediated via quantized relativistic fields \cite{Unruh,DeWitt}. In this case, quantum fields that interact with the UDWs propagate in flat \cite{QSignal,Jonsson2017,Rel:commun,Info:transm} or curved spacetime \cite{Comm:curv,Commun:BlackHole}, and constitute the \textit{quantum channel} between the two. The formalism employed to study these systems usually requires   perturbative approaches in order to obtain explicit solutions \cite{Cliche:Kempf2010,QSignal,Jonsson2017,Comm:curv,Commun:BlackHole}. Non-perturbative approaches have recently been explored with the help of gapless detectors \cite{Landulfo2016,Landulfo2021}, and in cases where qubit detectors  interact with the field very rapidly at a single instant of
time through a delta-like coupling \cite{Rel:commun,Info:transm,Tjoa2022}. 
 
In the present work, we study the channel capacity of the channel established between two particle detectors, modelled as harmonic oscillators \cite{Sciama,Hu:Matacz94,Bruschi2013,Martinez2013}, which are coupled via a massless scalar field in flat spacetime. The setup of two oscillators linearly interacting with a quantum field is formally equivalent to the Quantum Brownian Motion model found in the theory of open systems \cite{Hu:Paz:Zhang,Fleming:Hu:Roura,Hanggi,QLE1}. The time evolution of the reduced state of the oscillators admits an exact solution for all times allowing us to study the communication channel both for arbitrary detector-field coupling strengths and frequencies of the detectors. Furthermore, since harmonic oscillators are fundamentally Bosonic systems, there is an advantage compared to using qubits in communicating a classical message since one can arbitrarily increase the number of particles encoding the message. Consequently, bad performance of the quantum channel can be compensated by increasing the number of encoding Bosons \cite{Lupo_2011,Pilyavets_2012}.   

We quantify the reliability of the communication of classical messages using the classical capacity of the channel and we find that its functional dependence on time depends on the setup chosen for the detector systems. In the optimal setup, where the detectors turn on sharply, the communication between them has a long ``turning on period'' after which the capacity becomes nearly constant. The important aspect in this case is that this setup involves finite values for the detector couplings and frequencies such that it would be ``easy'' to reproduce them in a laboratory. Moreover, we also provide a strategy to decrease the ``turning on period'', with a cost in communication reliability. 

The paper is organized as follows. In Sec.~\ref{model} we introduce our model, by giving a short description of oscillator detectors and presenting a quantum Langevin equation that we employ to describe theirs dynamics. In Sec. \ref{commun} we build our communication protocol. In Sec.~\ref{capacities} we give a review on the classification of the quantum channels for Bosonic Gaussian systems and their capacity. In Sec.~\ref{Sec5} we study the transmissivity, noise and capacity of built quantum channel on a wide range of setups. Finally, in Sec.~\ref{conclusions} we summarise and discuss our main results.

\section{Quantum fields and oscillator detectors}
Here we provide an introduction to the formalisms necessary to our work.

Throughout this work we denote spatial vectors with boldface letters $(\xx)$, while spacetime vectors are represented by sans-serif characters $(\x)$. We use the signature $(+---)$ for the Minkowski spacetime metric. For the Fourier transform we employ the convention $\widetilde{f}(z)=\int_{-\infty}^{+\infty} dt\, e^{iz t}f(t)$, with the inverse Fourier transform $f(t)=(2\pi)^{-1}\int_{-\infty}^{+\infty}dz\, e^{-iz t}\widetilde{f}(z)$ respectively. Unless otherwise specified we set $\hbar=c=G_{\text{N}}=1$. We work in the Heisenberg picture.

\subsection{Harmonic oscillator detectors}\label{model}

We consider a massless scalar quantum field $\hat{\Phi}(\x)$ that propagates on a background (3+1)-dimensional Minkowski spacetime with metric $\eta_{\mu\nu}=\text{diag}(1,-1,-1,-1)$, see  \cite{Srednicki:2007}. The scalar field satisfies the Klein-Gordon equation $\square \hat{\Phi}(\x)=0$, where $\square=\eta^{\mu\nu}\partial_{\mu}\partial_{\nu}=\partial_t^2-\nabla^2$ is the d'Alembert operator \cite{Srednicki:2007}. The standard solutions to the Klein-Gordon equation are plane waves $\exp[-i k_\mu x^\mu]$, where $k^\mu\equiv(|\boldsymbol{k}|,\boldsymbol{k})$. The field can be obtained as a linear combination of such solutions and it reads
\begin{equation}\label{field}
    \hat{\Phi}(t,\xx)=\int\frac{d^3\kk}{\sqrt{(2\pi)^3 2|\kk|}}\left(\hat{a}_{\kk}e^{-i(|\kk|t-i\kk\cdot\xx)}+\text{H.c.}\right),
\end{equation}
where $\hat{a}_{\kk}$ and $\hat{a}_{\kk}^\dagger$ are the annihilation and creation operators of the plane wave with momentum $\kk$. They satisfy the  canonical commutation relations $[\hat{a}_{\kk},\hat{a}_{\kk'}^\dag]=\delta^{3}(\kk-\kk')$, while all others vanish \cite{Srednicki:2007}.

We next consider two static, non-interacting detectors labelled by $A$ and $B$, with unit masses $m_A=m_B=1$ and bare frequencies $\omega_\textrm{A}$ and $\omega_B$, that are placed within one-dimensional harmonic traps located at $\mathbf{x}_A$ and $\mathbf{x}_B$ respectively. Each detector is coupled to the field through the interaction Hamiltonian 
\begin{equation}\label{UDW:hamilt}
    \hat{H}_{\text{int}}(t)=\sum_{i=\{A,B\}}\lambda_{i}(t)\hat{x}_i(t)\otimes\hat\Phi_f(t,\mathbf{x}_i),
\end{equation}
where $\hat{x}_i$ is the displacement of each oscillator, $\lambda_{i}(t)$ describes how the coupling  between the $i$-th detector and the field is switched on and off, and we have introduced the spatial smeared field operator 
\begin{equation}
\hat\Phi_f(t,\mathbf{x}_i)=\int d^3\xx f_i(\xx-\xx_i)\hat\Phi(t,\xx),
\end{equation} 
where $f_i(\xx)$ is the so-called smearing function, and $\xx_i$ is the position of the center of mass of each detector \cite{HuLouko}.
The real-valued smearing functions $f_i(\xx)$ can be interpreted as the shape (thus providing the size) of each detector \cite{Schlicht_2004,Louko:Satz}. Note that by choosing a Dirac delta smearing $f(\xx)=\delta^3(\xx)$ the standard pointlike detector model is recovered. We will consider a sudden switching $\lambda_i(t)=\lambda_i\theta(t)$, where $\theta(x)$ is the Heaviside step function. In this way, the constants $\lambda_i$ identify the coupling strength between the $i$-th detector and the scalar field.

\subsection{The quantum Langevin equation}
The oscillator  detector  model characterized by the interaction Hamiltonian \eqref{UDW:hamilt} is a special case of the Caldeira-Leggett model of quantum Brownian motion \cite{CALDEIRA,breuer}. In this case, the scalar field plays the role of an environment characterized by an Ohmic spectral density. Working in the Heisenberg picture, the dynamics of the oscillators can be described by the quantum Langevin equation \cite{WeissQDS}, which reads
\begin{equation}\label{Langevin}
     \Ddot{\hat{x}}_i(t)+\omega_i^2\hat{x}_i(t)-\int_{0}^{t}ds\,\chi_{i}{}^j(t-s)\hat{x}_{j}(s)=\hat{\varphi}_i(t).
\end{equation}
Here, the repeated index $j$ is summed over $j=\{A,B\}$, the quantity $\hat{\varphi}_i(t):=\lambda_i\hat\Phi_f(t,\mathbf{x}_i)$ acts as an external force on each oscillator, and the matrix $\chi_{ij}=\chi_{i}{}^j$ defined by
\begin{equation}\label{disskernel}
    \chi_{ij}(t-t'):={\rm i}\theta(t-t')\expval{\big[\hat{\varphi}_i(t),\hat{\varphi}_{j}(t')\big]}
\end{equation}
is called the \emph{dissipation kernel} \cite{breuer}, which can be identified with the retarded propagator of the field \cite{birrell}.

Introducing the following vectors and matrix notation
\begin{equation}\label{matrix notation}
    x:=\begin{pmatrix}
\hat{x}_A \\
\hat{x}_B
\end{pmatrix}, \quad 
\mathbb{W}^2:=\begin{pmatrix}
\omega_\textrm{A}^2 & 0\\
0 &\omega_B^2
\end{pmatrix}, \quad 
\varphi:=\begin{pmatrix}
\hat\varphi_A \\
\hat\varphi_B
\end{pmatrix},
\end{equation}
we can recast the Langevin equation \eqref{Langevin} in the following compact matrix form
\begin{equation}\label{Langevin:matrix}
     \Ddot{x}(t)+\mathbb{W}^2x(t)-\int_{0}^{t}ds\,\chi(t-s)x(s)=\varphi(t).
\end{equation}
The solution of this equation reads
\begin{equation}\label{solution}
    x(t)=\dot{\mathbb{G}}(t)x(0)+\mathbb{G}(t)\dot{x}(0)+\int_{0}^{t}ds\,\mathbb{G}(t-s)\varphi(s),
\end{equation}
where $\mathbb{G}(t)=\left(G_{ij}\right)_{i,j=A,B}$  is the solution of the homogeneous part of Eq.~\eqref{Langevin:matrix} with initial conditions $\mathbb{G}(t\le0)=0$ (causality) and $\dot{\mathbb{G}}(0)=\mathds{1}$. It can be expressed through the Fourier transform
\begin{equation}\label{greenft}
    \widetilde{\mathbb{G}}(z)=\Big(-z^2\mathds{1}+\mathbb{W}^2-\widetilde{\chi}(z)\Big)^{-1},
\end{equation}
where $\widetilde{\chi}(z)$ is the Fourier transformed dissipation kernel.

\subsection{Gaussian state formalism}\label{Gaussian:formalism}
In this work we focus on Gaussian states of continuous variables systems. Gaussian states have Gaussian characteristic functions and are completely determined by their first and second moments \cite{Adesso_2014, RevModPhys.84.621,WANG20071}. Such states are paramount in quantum optics \cite{QO1,QO2}, where they can be used for quantum computing \cite{QComp1,QComp2} and sensing \cite{oh2020optimal}. 

Let us introduce the position operator $\hat{x}_i=1/\sqrt{2}\bigl(\hat{b}_i+\hat{b}_i^\dagger\bigr)$ and the momentum operator $\hat{p}_i=1/(\sqrt{2}{\rm i})\bigl(\hat{b}_i-\hat{b}_i^\dagger\bigr)$, where $i=\textrm{A},\textrm{B}$ labels the detector and $\hat{b}_i$ is the annihilation operator of the oscillator $i$ (not to be confused with the annihilation operator $a_\mathbf{k}$ of the scalar normal mode $\mathbf{k}$).
The first moment is defined as the vector
$\mathbf{d}:=(\langle \hat{x}_A\rangle,\langle \hat{p}_A\rangle,\langle \hat{x}_B\rangle,\langle \hat{p}_B\rangle)$,
where $\langle\cdot\rangle$ indicates the expectation value with respect to the detectors' state. More important to us is the \textit{covariance matrix} of seconds moments, defined by
\begin{equation}\label{cov:matrix}
    \sigma\coloneqq\begin{pmatrix}
\sigma_{xx} & \sigma_{xp}\\
\sigma_{px} & \sigma_{pp}
\end{pmatrix},
\end{equation}
with $\sigma_{\alpha\beta}(t)\coloneqq\frac{1}{2}\bigl\langle\bigl\{ \hat{\alpha}(t),\hat{\beta}^T(t)) \bigr\}\bigr\rangle-\left\langle\hat{\alpha}(t)\right\rangle\langle\hat{\beta}^T(t)\rangle$ ,
where $\alpha,\beta\in\left\{x,p\right\}$.
Since the relevant (i.e., entropic) quantities we are interested in do not depend on the first moments, from now on we focus exclusively on the covariance matrix \eqref{cov:matrix}. It is worth noticing that, by exchanging the second and third rows and columns of the covariance matrix \eqref{cov:matrix}, we can rewrite it in the form
\begin{equation}\label{covmatrixcorrbasis}
    \mathbf{\sigma}=\left(\begin{matrix}
    \sigma_{\textrm{AA}}&\sigma_{\textrm{AB}}\\\sigma_{\textrm{BA}}&\sigma_{\textrm{BB}}
    \end{matrix}\right).
\end{equation}
In the latter case, the covariance matrix $\sigma_{\textrm{AA}}$ ($\sigma_{\textrm{BB}}$) represents exactly the state of the detector $A$ ($B$). The matrix $\sigma_{\textrm{AB}}=\sigma_{\textrm{BA}}^T$ instead describes the correlation between the detectors \cite{PhysRevA.69.022318}.

\section{Communication protocol}\label{commun}
Our aim is to study how information encoded into states of the detector $A$ (held by Alice) can be faithfully transmitted to the detector $B$ (held by Bob). To this end, since the whole system is composed by harmonic oscillators, we consider the two detectors to be prepared initially in a separable two-mode Gaussian state.

In the communication protocol considered here, Alice prepares the detector $A$ in a state that is sent to Bob through the quantum field by means of the detector-field interaction activated at $t=0$. We want to know how reliably the signal is transmitted to Bob as a function of time $t$. The fact that there is no detector-field interaction before $t=0$ ensures that the detectors are completely uncorrelated at $t=0$, so that $\sigma_{\textrm{AB}}(0)=0$. 

We also assume that the detectors and the field are initially prepared in a separable state. The time evolution of the expectation value of the operators $\hat{x}_{i=\textrm{A},\textrm{B}}$ is given by Eq.~\eqref{solution}. Since we work in the Heisenberg picture and we have chosen the detectors to have unit mass, the time evolution of the momentum $\hat{p}_i$ of the detector $i$ reads  $\hat{p}_i(t)=\dot{\hat{x}}_i(t)$. Finally, using Eq.~\eqref{solution} and its derivative, we can compute the time evolution of the elements of the covariance matrix \eqref{cov:matrix}, and we find
\begin{align}\label{sigmaxx}
    \sigma_{xx}(t)=&\dot{\mathbb{G}}(t)\sigma_{xx}(0)\dot{\mathbb{G}}^T(t)+\mathbb{G}(t)\sigma_{pp}(0)\mathbb{G}^T(t)\nonumber\\&+\dot{\mathbb{G}}(t)\sigma_{xp}(0)\mathbb{G}^T(t)+\mathbb{G}(t)\sigma_{px}(0)\dot{\mathbb{G}}^T(t)\nonumber\\&+\int_{0}^{t}ds\int_{0}^{t}ds'\mathbb{G}(t-s)\nu(s,s')\mathbb{G}^T(t-s'),
\end{align}
\begin{align}\label{sigmaxp}
    \sigma_{xp}(t)=&\dot{\mathbb{G}}(t)\sigma_{xx}(0) \Ddot{\mathbb{G}}^T(t)+\mathbb{G}(t)\sigma_{pp}(0)\dot{\mathbb{G}}^T(t)\nonumber\\&+\dot{\mathbb{G}}(t)\sigma_{xp}(0)\dot{\mathbb{G}}^T(t)+\mathbb{G}(t)\sigma_{px}(0) \Ddot{\mathbb{G}}^T(t)\nonumber\\&+ \int_{0}^{t}ds\int_{0}^{t}ds'\mathbb{G}(t-s)\nu(s,s')\dot{\mathbb{G}}^T(t-s'),
\end{align}
\begin{align}\label{sigmapp}
    \sigma_{pp}(t)=& \Ddot{\mathbb{G}}(t)\sigma_{xx}(0) \Ddot{\mathbb{G}}^T(t)+\dot{\mathbb{G}}(t)\sigma_{pp}(0)\dot{\mathbb{G}}^T(t)+\nonumber\\&\Ddot{\mathbb{G}}(t)\sigma_{xp}(0)\dot{\mathbb{G}}^T(t)+\dot{\mathbb{G}}(t)\sigma_{px}(0) \Ddot{\mathbb{G}}^T(t)\nonumber\\& +\int_{0}^{t}ds\int_{0}^{t}ds'\dot{\mathbb{G}}(t-s)\nu(s,s')\dot{\mathbb{G}}^T(t-s').
\end{align}
Here we have introduced the quantity, 
\begin{equation}\label{noisekernel}
    \nu(t,t'):=\frac{1}{2}\expval{\left\{\hat\varphi(t),\hat\varphi^T(t')\right\}},
\end{equation}
known as the \emph{noise kernel} \cite{breuer},  which can be identified with the Hadamard function of the field \cite{birrell}. 

We note that the noise kernel combined with the dissipation kernel provide the Wightman two-point correlation function of the field, namely
$ \mathcal{W}(\tau,\tau')=\expval{\hat\varphi(\tau)\hat\varphi(\tau')}\nonumber\equiv \nu(\tau,\tau')+i\chi(\tau,\tau')$. When the state of the field is stationary and the detectors follow a stationary trajectory \cite{Letaw}--as it is the case of static detectors in Minkowski spacetime--the Wightman function depends only on the difference $\tau-\tau'$ and we can write $\mathcal{W}(\tau,\tau')=\mathcal{W}(\tau-\tau')$.

In our communication protocol, the \emph{input} state is characterized by the covariance matrix $\sigma_{in}:=\sigma_{\textrm{AA}}(0)$ of the detector $A$ at $t=0$ while the \emph{output} state by the covariance matrix $\sigma_{out}:=\sigma_{\textrm{BB}}(t)$ of the detector $B$ at a certain time $t>0$. 
In order to obtain $\sigma_{\textrm{BB}}(t)$ we write the covariance matrix \eqref{cov:matrix} at the time $t$ into the form \eqref{covmatrixcorrbasis}. Using the fact that $\sigma_{\textrm{AB}}(0)=0$, we find that the covariance matrix of the subsystem of the detector $B$ at time $t$ is given by
\begin{align}\label{explicit output}
    \sigma_{out}&=\sigma_{\textrm{BB}}(t)= T_{\textrm{BB}}\sigma_{\textrm{BB}}(0)T_{\textrm{BB}}^T(t)+T_{\textrm{BA}}(t)\sigma_{in}T_{\textrm{BA}}^T(t)\nonumber\\&+\int_0^tds\int_0^tds'\eta(t-s)\nu(s,s')\eta^T(t-s'),
\end{align}
where the matrices $T_{ij}$, with $i,j\in\{A,B\}$, and $\eta$ are defined respectively as
\begin{equation}\label{eta}
    T_{ij}:=
    \left(\begin{matrix}
    \dot{G}_{ij}&G_{ij}\\\ddot{G}_{ij}&\dot{G}_{ij}
    \end{matrix}\right),
\quad
    \eta:=
    \left(\begin{matrix}
    G_{\textrm{BA}}&G_{\textrm{BB}}\\\dot{G}_{\textrm{BA}}&\dot{G}_{\textrm{BB}}
    \end{matrix}\right).
\end{equation}

\section{Gaussian channels and capacities}\label{capacities}
The input-to-output transformation of Eq.~\eqref{explicit output} realizes a one-mode Gaussian channel. For such kind of channels, the relation between the input and the output covariance matrices is of the form
\begin{equation}\label{interestingevo}
    \sigma_{out}=\mathbb{T}\sigma_{in}\mathbb{T}^T+\mathbb{N},
\end{equation}
where $\mathbb{T}$ and $\mathbb{N}$ are $2\times2$ matrices expressing the transmissivity and noisy properties of the channel respectively \cite{Caruso2006a}. 
Analyzing Eq.~\eqref{explicit output} we find that  
\begin{equation}\label{Tmatrix}
    \mathbb{T}(t)=T_{\textrm{BA}}(t)=\left(\begin{matrix}
    \dot{G}_{\textrm{BA}}(t)&G_{\textrm{BA}}(t)\\\ddot{G}_{\textrm{BA}}(t)&\dot{G}_{\textrm{BA}}(t)
    \end{matrix}\right),
\end{equation}
and
\begin{align}\label{noisematrix}
    \mathbb{N}(t)=&T_{\textrm{BB}}(t)\sigma_{\textrm{BB}}(0)T_{\textrm{BB}}(t)^T\nonumber\\&\quad+\int_0^tds\int_0^tds'\eta(t-s)\nu(s,s')\eta(t-s').
\end{align}
These quantities are key to our analysis below.

\subsection{Channel classification}
In general, the entropic quantities that can be computed for a one-mode Gaussian channel are characterized by the aforementioned $2\times2$ matrices $\mathbb{T}$ and $\mathbb{N}$. These matrices can be reduced to the so-called \textit{canonical form} \cite{WilliamsonT,Caruso2006} by applying a symplectic transformation $S_A$ on the input covariance matrix (called \textit{pre-processing} transformation) and another symplectic transformation on the output covariance matrix $S_B$ (called \textit{post-processing} transformation).\footnote{In some singular cases, the matrices $\mathbb{T}$ and $\mathbb{N}$ have rank 1 and the expressions \eqref{Tcanonical:Ncanonical} are not valid. However, we shall not consider these cases here.} The canonical form $\mathbb{T}_{\text{c}}$ and $\mathbb{N}_{\text{c}}$ of the matrices $\mathbb{T}$ and $\mathbb{N}$ reads
\begin{align}\label{Tcanonical:Ncanonical}
\mathbb{T}_{\text{c}}&=S_A\mathbb{T}S_B^T=\sqrt{|\tau|}\mathds{1},\nonumber\\   
\mathbb{N}_{\text{c}}&=S_B\mathbb{N}S_B^T=\sqrt{W}\mathds{1},
\end{align}
where $\tau$ is real and $W\in[1/4,+\infty)$.
The pre-processing and post-processing matrix can be explicitly derived in terms of the elements of the matrices $\mathbb{T}$ and $\mathbb{N}$ as
\begin{equation}\label{preprocMatrix}
    S_A=\frac{\sqrt[4]{W}}{\sqrt{N_{11}\tau}}\left(\begin{matrix}
        \frac{N_{11}T_{22}-N_{12}T_{12}}{\sqrt{W}}&-T_{12}\\
        \frac{N_{12}T_{11}-N_{11}T_{21}}{\sqrt{W}}&T_{11}
    \end{matrix}\right),
\end{equation}
\begin{equation}
    S_B=\frac{\sqrt[4]{W}}{\sqrt{N_{11}}}\left(\begin{matrix}
        1&0\\
        -\frac{N_{12}}{\sqrt{W}}&\frac{N_{11}}{\sqrt{W}}
    \end{matrix}\right).
\end{equation}
It is immediate to see that det$(S_{B})=1$ and det$(S_A)=\text{sign}(\tau)$, while computing the determinant in Eqs.~\eqref{Tcanonical:Ncanonical} we have $\tau=\det(\mathbb{T})$ and $W=\det(\mathbb{N})$. In other words, the parameter $\tau$ can be regarded as the portion of input signal transmitted to the output. Then, the one-mode Gaussian channels are classified depending on their value of $\tau$, see \cite{Caruso2006}. We have:
\begin{itemize}
    \item $\tau>1$: An \textit{amplifier channel}; 
    \item $\tau=1$:  An \textit{additive noise channel}; 
    \item $0<\tau<1$: A \textit{lossy channel};
    \item $\tau=0$: An \textit{erasure channel};
    \item $\tau<0$: A \textit{conjugate channel} to the amplifier one. 
\end{itemize}
From the determinant of $\mathbb{N}$, i.e. $W$, we can evaluate the average additive classical noise $\overline{n}$ produced by the channel. In particular, for the class of additive noise channels we have
\begin{equation}\label{addBnoise}
    \overline{n}=\sqrt{W}.
\end{equation}
Instead, for all the other classes, we have
\begin{equation}\label{additivenoise}
    \overline{n}=\frac{\sqrt{W}}{\left|1-\tau\right|}-\frac{1}{2}.
\end{equation}

\subsection{Classical capacity}\label{ssecCapacities}
An ideal one-mode Gaussian channel would be a channel in which the output is identical to the input, namely $\tau=1$ and $\overline{n}=0$. Deviation from any of these conditions gives a noisy contribution to the channel, compromising its transmission capability. To know how well a channel transmits information, one has to study a quantity which takes into account both of the aforementioned contributes. Such a quantity is the \textit{capacity} of the quantum channel \cite{Hausladen1996,Wolf_2005,Gyongyosi_2018}. The \textit{classical capacity} (\textit{quantum capacity}) of a quantum channel is identified as the maximum rate of classical (quantum) information that the channel can transmit reliably.\footnote{A formal definition of \textit{reliable transmission} can be found, e.g., in \cite{Hausladen1996}.}

The classical capacity is obtained, in general, by maximizing the \textit{Holevo information} over all the possible input states \cite{Holevo1973BoundsFT,Holevo1999}. In the following, to avoid the regularization problem  
 we restrict our input states to Gaussian states over each channel use \cite{Devetak2003}. As a consequence, our result for the classical capacity has to be intended as a lower bound of it.

In the considered protocol Alice, in order to encode her classical message, starts from a state $\sigma_{in}$ and then performs a displacement  according to a Gaussian distribution with covariance $\sigma_{enc}$. The quantum channel is then the Gaussian map
${\cal N}$ mapping $ (\sigma_{in}+\sigma_{enc}) \mapsto \sigma_{out}$,
where $\sigma_{out}$ is the covariance matrix of the Gaussian states received by Bob at the detector $B$. The Holevo information $\mathcal{X}$ relative to this protocol has already been computed in the literature  \cite{Lupo_2011,Pilyavets_2012}, and it reads
\begin{align}\label{HolevoInfGeneral}
    \mathcal{X}(\sigma_{in},\sigma_{enc},\mathcal{N}):=&{\sf S}\left(\mathbb{T}(\sigma_{in}+\sigma_{enc})\mathbb{T}^T+\mathbb{N}\right)\nonumber\\
    &-{\sf S}(\mathbb{T}\sigma_{in}\mathbb{T}^T+
    \mathbb{N}),
\end{align}
where ${\sf S}$ is the Von Neumann entropy of the state represented by the covariance matrix $\sigma$. When using covariance matrices the Von Neumann entropy has the simple expression ${\sf S}(\sigma)=h_+(\nu)-h_-(\nu)$, where $\nu$ is the symplectic eigenvalue of the matrix $\sigma$ and $h_\pm:=(\nu\pm1/2)\log(\nu\pm1/2)$. When $\sigma$ is a $2\times2$ matrix, its symplectic eigenvalue coincides with the square root of its determinant.
 Therefore, the lower bound to the classical capacity of the channel $\mathcal{N}$ is 
\begin{equation}
C(\mathcal{N})=\max_{\sigma_{in},\sigma_{enc}}\mathcal{X}(\sigma_{in},\sigma_{enc},\mathcal{N}).
\end{equation}
By writing the matrices $\mathbb{T}$ and $\mathbb{N}$ in their canonical form \eqref{Tcanonical:Ncanonical} (and reminding that the post processing $S_B$ does not change the Von Neumann entropy), the Holevo information \eqref{HolevoInfGeneral} becomes $\mathcal{X}={\sf S}(\mathbb{T}_{\text{c}}S_A(\sigma_{in}+\sigma_{enc})S_A^T\mathbb{T}_{\text{c}}+\mathbb{N}_{\text{c}})-{\sf S}(\mathbb{T}_{\text{c}}S_A\sigma_{in}S_A^T\mathbb{T}_{\text{c}}+\mathbb{N}_{\text{c}})$. 
We can calculate and maximize analytically the Holevo information in this form by applying a Bloch-Messiah decomposition to decompose the pre-processing matrix $S_A$ (defined in Eq.~\eqref{preprocMatrix}), see \cite{Braunstein_2005}. In particular, this means that we can write $S_A=\mathbb{R}\mathbb{D}\mathbb{R}'$, where $\mathbb{R}$ and $\mathbb{R}'$ are orthogonal matrices and $\mathbb{D}=\text{diag}(r^{1/2},r^{-1/2})$ is a squeezing matrix. It is possible to calculate $r$ from Eq.~\eqref{preprocMatrix}, leading to the result
\begin{equation}\label{preprocessing}
    r=\frac{1}{2}(\mathcal{T}\pm\sqrt{\mathcal{T}^2-4}),
\end{equation}
where we have defined 
\begin{align}
    \mathcal{T}:=&\frac{N_{22}(T_{11}^2+T_{12}^2)+N_{11}(T_{21}^2+T_{22}^2)}{\sqrt{W}|\tau|}\notag\\
    &-\frac{2N_{12}(T_{11}T_{21}+T_{12}T_{22})}{\sqrt{W}|\tau|}.
\end{align}
Then, the matrix $\mathbb{R}'$ can be absorbed into the matrices $\sigma_{in}$ and $\sigma_{out}$. At this point, the Holevo information becomes:
\begin{equation}\label{HolInfnonFinal}
    \mathcal{X}={\sf S}(|\tau| \mathbb{R}\mathbb{D}(\sigma_{in}+\sigma_{enc})\mathbb{D}\mathbb{R}^T+\mathbb{N}_{\text{c}})-{\sf S}(|\tau|\mathbb{R}\mathbb{D}\sigma_{in}\mathbb{D}\mathbb{R}^T+\mathbb{N}_{\text{c}}).
\end{equation}
Since $\mathbb{N}_c=\mathbb{R}\mathbb{N}_c\mathbb{R}^T$ because $\mathbb{N}_c\propto\mathds{1}$ in its canonical form, $\mathbb{R}$ becomes an orthogonal transformation acting on the argument of the entropy ${\sf S}$. However, by its definition, ${\sf S}$ is unaffected by orthogonal transformations. For this reason, the matrix $\mathbb{R}$ can be neglected and the Holevo information becomes  (see also Ref.~\cite{Lupo_2011}):
\begin{equation}\label{HolInfFinal}
    \mathcal{X}={\sf S}(|\tau| \mathbb{D}(\sigma_{in}+\sigma_{enc})\mathbb{D}+\mathbb{N}_{\text{c}})-{\sf S}(|\tau|\mathbb{D}\sigma_{in}\mathbb{D}+\mathbb{N}_{\text{c}}).
\end{equation}

For Alice it would be optimal to encode the message into a state with covariance matrix $\sigma_{enc}$ whose symplectic eigenvalues are as large as possible. In this way, the classical capacity would always be infinite. 
To avoid this unphysical situation it is customary to set a bound $E$ on the average energy she can use (see e.g.~\cite{giovannetti2015solution}). Explicitly, the bound reads
\begin{equation}\label{Energyconstraint}
\frac{1}{2}\text{Tr}(\mathbb{D}(\sigma_{in}+\sigma_{enc})\mathbb{D})\le \frac{E}{\omega_\textrm{A}}
\equiv \frac{1}{2}+\bar{N},
\end{equation}
where $\bar{N}$ represents the average number of particles used to encode the message.

From the \textit{input purity theorem} \cite{Pilyavets_2012}, a pure input state maximizes the Holevo information. A generic pure input state can be written as $\sigma_{in}=1/2\,\text{diag}(j,j^{-1})$, with $j>0$.
 For the encoding, we write $\sigma_{enc}=\text{diag}(x,y)$ with $x,y>0$. For simplicity, we define $J:=jr$, $X:=xr$ and $Y:=y/r$. Using them the Holevo information \eqref{HolInfFinal} becomes
{\small
\begin{align}\label{linearHolInf}
    \mathcal{X}(J,X,Y)=&h\left(\frac{1}{2}\sqrt{\left(\tau J+2\sqrt{W}+2 \tau X\right)\left(\frac{\tau}{J}+2\sqrt{W}+2 \tau Y\right)}\right)\nonumber\\
    &-h\left(\frac{1}{2}\sqrt{\left(\tau J+2\sqrt{W}\right)\left(\frac{\tau}{ J}+2\sqrt{W}\right)}\right).
\end{align}
}
The aim is to maximize $\mathcal{X}(J,X,Y)$ over $X$, $Y$ and $J$ that satisfy the conditions
\begin{equation}\label{conditions}
    \begin{cases}
        \frac{J}{2}+X+\frac{1}{2J}+Y\le 2\frac{E}{\omega_\textrm{A}},\\
        X>0,\\
        Y>0.
    \end{cases}
\end{equation}
The first condition in \eqref{conditions} comes from the condition \eqref{Energyconstraint}. Since we want to use as much encoding energy as possible, we impose the equality in the first condition \eqref{conditions}. In this way, we can express $Y$ in terms of $X$ and $J$, and we find
\begin{equation}\label{y}
    Y=2\frac{E}{\omega_\textrm{A}}-\frac{1}{2J}-\frac{J}{2}+X\,.
\end{equation}
With this new relation we simplify Eq.~\eqref{linearHolInf}. We then note that, since $X$ is present only in the first term of Eq.~\eqref{linearHolInf} and  $h(x)$ is a monotonic function for $x$, we can find analytically the optimal $X$ for the Holevo information \eqref{linearHolInf}. Imposing the second and third conditions in Eq.~\eqref{conditions}, we get\footnote{To be rigorous, $X=0$ in Eq.~\eqref{xmax} has to be intended as the limit $X\to0^+$. Indeed, since $\sigma_{enc}$ is definite positive, $X=0$ should not be allowed.}
\begin{align}\label{xmax}
    X=
    \left\{\begin{matrix*}
    0, &\text{if}&  2\frac{E}{\omega_\textrm{A}}<J<2\frac{E}{\omega_\textrm{A}}+\sqrt{4\frac{E^2}{\omega_\textrm{A}^2}-1}\,\\
        \frac{E}{\omega_\textrm{A}}-\frac{J}{2}, &\text{if}& \frac{\omega_\textrm{A}}{2E}<J<2\frac{E}{\omega_\textrm{A}}.
        \end{matrix*}
        \right.
\end{align}
It is interesting to notice how, by fixing the encoding energy $E$, the conditions \eqref{conditions} limit the possible values of $J$ between $1/(2E)$ to $2E+\sqrt{4E^2-1}$. In other words, by fixing an encoding energy the number of states that one can create with this encoding energy is limited. As expected, the range of possible $J$ increases as a function of $E$ because we clearly have an increasing freedom of choosing different input states. Conversely, if $E=\frac{\omega_\textrm{A}}{2}$, i.e., the minimum possible encoding energy (vacuum energy), the only choices for our parameters are $J=1$, $X=0$ and $Y=0$, respectively\footnote{To compare this result with the literature (see e.g. Ref.~\cite{Pilyavets_2012}), the range in which $X\ne0$ ($X=0$), from Eq.~\eqref{xmax}, is called \textit{third stage} (\textit{second stage}).}.

Inserting Eq.~\eqref{xmax} into Eq.~\eqref{linearHolInf}, we get an expression for the Holevo information dependent only on $J$. The optimization of the latter can be performed numerically, giving the lower bound for the classical capacity.

\subsection{Quantum capacity}\label{quantumC}
We spend few words here on the quantum capacity of a one-mode Gaussian channel. We leave the reader to the literature for further information \cite{Caruso2006,HolevoQC,Br_dler_2015}.

Again, to avoid the regularization problem \cite{Devetak2003}, we constraint our input states into states which are separable over each channel use. Therefore the quantity of interest becomes the single letter version $Q^{(1)}$ of the quantum capacity.
We recall that this capacity represents a lower bound for the true quantum capacity and is obtained by maximizing the so-called \textit{coherent information} $I_{c}$ defined as the difference between the von Neumann entropy of state
resulting after the application of the channel and the von Neumann entropy of the state resulting after the application of the complementary channel.
The maximization of this quantity is again achieved when the encoding energy $E$ is infinite. However, this time, there is no need to put an energy constraint since $Q^{(1)}$ remains finite in the limit $E\rightarrow\infty$. In this limit, as long as $\tau$ is positive and different from $1$, the value of the coherent information is 
\begin{equation}\label{maxcoherentinf2}
    I_\textrm{c}(E\to\infty)=\log\frac{\tau}{|1-\tau|}-h\left(\frac{\sqrt{W}}{|1-\tau|}\right).
\end{equation}
The single letter formula for the quantum capacity is hence
\begin{equation}\label{quantumcapacity}
    Q^{(1)}=\max(0,I_\textrm{c}(E\to\infty)).
\end{equation}
From Eqs.~\eqref{maxcoherentinf2} and \eqref{quantumcapacity}, we can notice that $\tau<1/2$ implies $Q^{(1)}=0$, confirming the validity of the no-cloning theorem \cite{NoCloning}.
Furthermore, we also note that if $\tau$ is negative, it is possible to show that $Q^{(1)}$ is always zero \cite{HolevoQC}. 

Manipulating Eq.~\eqref{maxcoherentinf2} and using Eq.~\eqref{additivenoise}, in the limit $\tau\to1$ the maximized coherent information converges to
\begin{equation}\label{BmaximizedcoherentInf}
    I_\textrm{c}(E\to\infty)\overset{\tau\to1}{\longrightarrow}\log\left(\frac{1}{\sqrt{W}}\right).
\end{equation}

\section{Quantum channel features}\label{Sec5}
We consider that both the detectors are described by the same Gaussian spatial profile
\begin{equation}\label{gaussian shape}
    f_i(\xx-\xx_i)=\frac{1}{(\sqrt{\pi}\sigma)^3}e^{-(\xx-\xx_i)^2/\sigma^2},
\end{equation}
where $\sigma$ determines the effective size of each detector. The communication properties of the protocol introduced in Sec.~\ref{commun} will be studied for the regime $\sigma\ll d$, i.e., when the effective size of the detectors is negligible with respect to their distance $d$. This situation is relevant, for example, when one considers  communication between satellites and quantum probes in the outer space that are usually placed very far from each other.\footnote{This scenario is also realistic in communication protocols that take into account spacetime curvature, which usually manifests itself on large scales, and thus, becomes important when the detectors are very distant.} In this case, the detectors can be considered as point-like objects implying that
\begin{equation}\label{approximation}
    f_i(\xx-\xx_i)=\frac{1}{(\sqrt{\pi}\sigma)^3}e^{-(\xx-\xx_i)^2/\sigma^2}\sim\delta^3(\mathbf{x}-\mathbf{x}_i)
\end{equation}
effectively.
In this limit, the elements of the dissipation kernel read\footnote{The dissipation kernel matrix $\chi(t)$ and the noise kernel matrix $\nu(t)$ are analytically reported in Appendix \ref{NewAppendixA} for Gaussian detectors}
\begin{equation}\label{dissKernelOffD}
    \chi_{\textrm{AB}}(t)=\chi_{\textrm{BA}}(t)=-\frac{\lambda_\textrm{A}\lambda_\textrm{B}}{4\pi d}\theta(t)\Big(\delta(t+d)-\delta(t-d)\Big),
\end{equation}
\begin{equation}\label{DissKernelD}
    \chi_{KK}(t)=-\frac{\lambda^2_i}{2\pi}\theta(t)\delta'(t).
\end{equation}
A similar calculation can be performed for the noise kernel matrix $\nu(t)$. The off-diagonal elements read
\begin{equation}
    \label{convergentkernel}
    \nu_{\textrm{AB}}(t)=\nu_{\textrm{BA}}(t)=-\frac{1}{8\pi^2d}\mathcal{P}\left(\frac{1}{t-d}-\frac{1}{t+d}\right)\,,
\end{equation}
where $\mathcal{P}$ denotes the Cauchy principal value, while the diagonal elements are
\begin{equation}\label{kerneldiag}
    \nu_{KK}(t)=-\frac{1}{4\pi^2 }\mathcal{P}\left(\frac{1}{t^2}\right)\,.
\end{equation}
We proceed by calculating the Green function solution $\mathbb{G}(t)$ to the homogeneous Langevin equation
\begin{equation}\label{HomLangevin}
    \ddot{\mathbb{G}}(t)+\Omega^2\mathbb{G}(t)-\int_0^t \chi(t-s)\mathbb{G}(s)ds=\mathds{1}\delta(t)\,,
\end{equation}
with initial conditions $\mathbb{G}(t\le0)=0$ and $\dot{\mathbb{G}}(0^+)=\mathds{1}$. By inserting the dissipation kernel elements \eqref{dissKernelOffD} and \eqref{DissKernelD} into Eq.~\eqref{HomLangevin}, and introducing a frequency cut-off $\omega_c$ by setting $\delta(0)=\frac{1}{\sqrt{2\pi}\sigma}=\omega_c$, see \cite{breuer},
the homogeneous Langevin equation reduces to the following system of differential equations
{\footnotesize
    \begin{align}\label{systemLangevin}
    \ddot{G}_{\textrm{AA}}(t)-\Sigma_\textrm{A}^2G_{\textrm{AA}}(t)+2\gamma_\textrm{A}\dot{G}_{\textrm{AA}}(t)=&2\frac{\sqrt{\gamma_\textrm{A}\gamma_\textrm{B}}}{d}G_{\textrm{AB}}(t-d)\theta(t-d)\,;\nonumber\\
    \ddot{G}_{\textrm{AB}}(t)-\Sigma_\textrm{A}^2G_{\textrm{AB}}(t)+2\gamma_\textrm{A}\dot{G}_{\textrm{AB}}(t)=&2\frac{\sqrt{\gamma_\textrm{A}\gamma_\textrm{B}}}{d}G_{\textrm{BB}}(t-d)\theta(t-d)\,;\nonumber\\
    \ddot{G}_{\textrm{BA}}(t)-\Sigma_\textrm{B}^2G_{\textrm{BA}}(t)+2\gamma_\textrm{B}\dot{G}_{\textrm{BA}}(t)=&2\frac{\sqrt{\gamma_\textrm{A}\gamma_\textrm{B}}}{d}G_{\textrm{AA}}(t-d)\theta(t-d)\,;\nonumber\\
    \ddot{G}_{\textrm{BB}}(t)-\Sigma_\textrm{B}^2G_{\textrm{BB}}(t)+2\gamma_\textrm{B}\dot{G}_{\textrm{BB}}(t)=&2\frac{\sqrt{\gamma_\textrm{A}\gamma_\textrm{B}}}{d}G_{\textrm{BA}}(t-d)\theta(t-d)\,,
    \end{align}
}
where $\gamma_i\coloneqq\frac{\lambda_i^2}{8\pi}$ plays the role of the field-detector coupling and we have defined $\Sigma_i^2\coloneqq\sqrt{\frac{8}{\pi}}\frac{\gamma_i}{\sigma}-\omega^2$, for $i=$A,B.

An analytical solution to the above system of differential equations is provided in Eqs.~\eqref{GreenAA1st} and \eqref{greenABfirst:appendix} in Appendix \ref{appendixA}, and it gives us the expression of the Green functions solutions $G_{ij}(t)$ for $i,j=$A,B. These solutions have been obtained applying either one of the following conditions:
\begin{itemize}
    \item[\textbf{C1}.] $|\Sigma_i^2|\gg2\sqrt{\gamma_{\textrm{A}}\gamma_{\textrm{B}}}/d$ with $i=\textrm{A},\textrm{B}$.
    \item[\textbf{C2}.] $0\le t <2d$.
\end{itemize}

Therefore, we choose to study the communication protocol analytically at all times $t$ only when $|\Sigma_i^2|\gg2\sqrt{\gamma_i\gamma_j}/d$, providing numerical results when the latter is not satisfied. 

Before continuing, it is worth stating that, by numerically computing the parameter $r$ in Eq.~\eqref{preprocessing}, we find that $r\sim1$ with an error $\sim10^{-8}$ in all the cases later described. In this way, we can consider by now $r=1$ where, as reported in the literature \cite{Lupo_2011}, the optimization of the Holevo information \eqref{HolInfFinal} occurs when $X=Y=E/\omega_\textrm{A}$ and when $J=1$, independently from the encoding energy $E$. As a consequence, the lower bound for the classical capacity $C$ is given exactly by the following equation
\begin{equation}\label{CapacitySimplified}
    C=h\left(\tau\frac{E}{\omega_\textrm{A}}+\sqrt{W}\right)-h\left(\frac{\tau}{2}+\sqrt{W}\right).
\end{equation}

\subsection{Identical detectors}
We start by studying the simplified case where the two detectors are identical; thus, $\gamma\coloneqq\gamma_\textrm{A}=\gamma_\textrm{B}$ and $\omega\coloneqq\omega_\textrm{A}=\omega_B$. As a consequence we also have $\Sigma^2=\Sigma_\textrm{A}^2=\Sigma_\textrm{B}^2$.
We will divide our analysis according to the different cases that arise due to the condition \textbf{C1}.

\subsubsection{Case I: $|\Sigma^2|\gg2\gamma/d$ with $\Sigma^2<0$}
Here we provide the results when the condition $|\Sigma^2|\gg2\gamma/d$ is satisfied. The Green function $G_{\textrm{AB}}(t)$ is provided in Eq.~\eqref{greenABfirst:appendix}, reducing to Eq.~\eqref{eq: transmission dec equals} when the detectors are identical. The transmissivity $\tau$ can be computed using Eqs.~\eqref{Tmatrix} and \eqref{Tcanonical:Ncanonical}, giving:
{\small
\begin{align}\label{eq: transm dec equals}
    \tau=&\frac{\gamma^2}{d^2}\frac{(t-d)^2e^{-2\gamma (t-d)}}{\gamma_\Sigma^2}\left(\frac{\sinh^2\left(\gamma_\Sigma(t-d)\right)}{\gamma_\Sigma^2(t-d)^2}-1\right)\,.
\end{align}
}
Here we have introduced $\gamma_\Sigma:=\sqrt{\gamma^2+\Sigma^2}$ for simplicity of presentation.

In Appendix \ref{appendixB} we compute the noise for this case. When the detectors are identical, and when the condition $|\Sigma^2|\gg2\gamma/d$ holds, an analytic expression of the noise is provided in Eq.~\eqref{detN}. Furthemore, once we obtain $\tau$ and $W$, we can compute the lower bound of the classical capacity with Eq.~\eqref{CapacitySimplified}, imposing an encoding energy $E$.

\begin{figure}[ht!]
    \centering
    \includegraphics[width=0.9\linewidth]{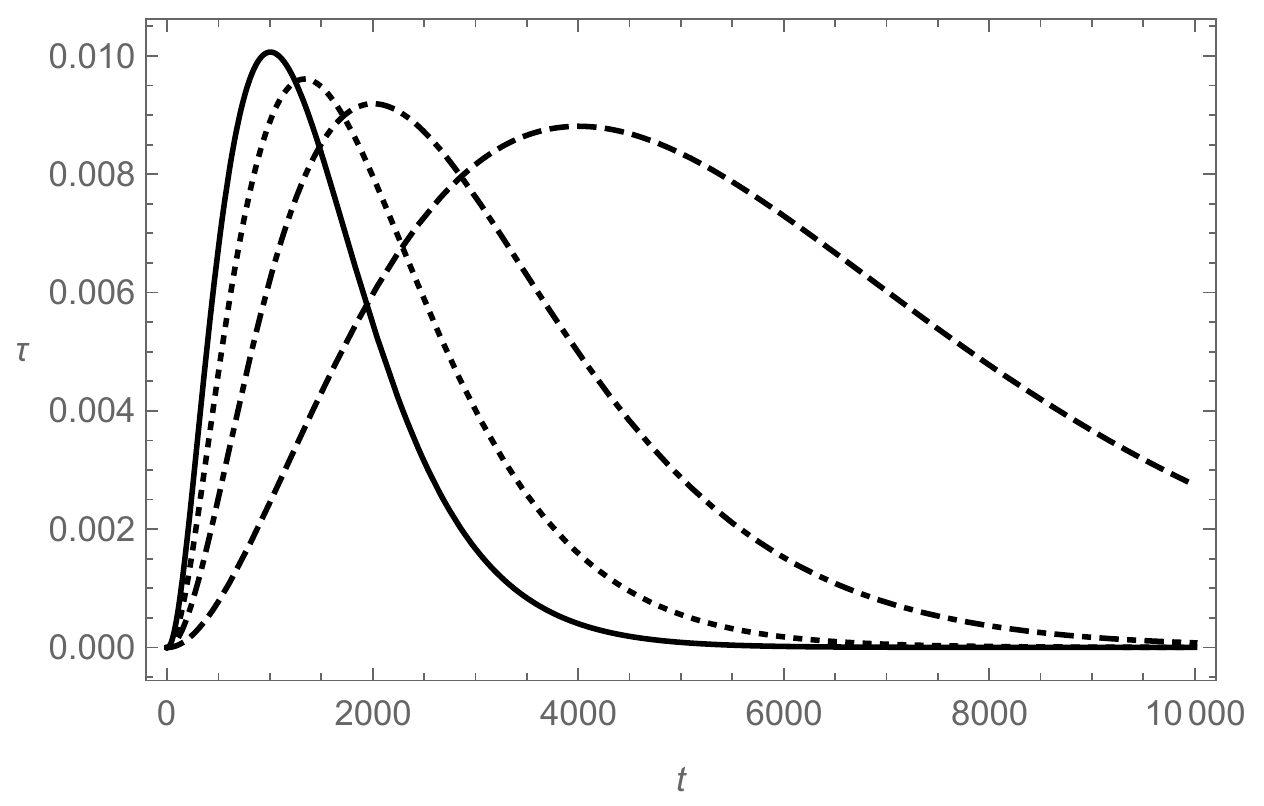}
    \caption{Transmissivity $\tau$ as function of time $t$ from Eq.~\eqref{eq: transm dec equals} for different values of the coupling $\gamma$, when the parameter $\Sigma^2$ is negative. In particular $\gamma=1\cdot10^{-3}$ (thick line), $\gamma=0.75\cdot10^{-3}$ (dotted line), $\gamma=0.5\cdot10^{-3}$ (dot-dashed line), $\gamma=0.25\cdot10^{-3}$ (dashed line). The other parameters used are $\sigma=0.01$, $\omega=1$, $d=4$.}
    \label{tau1}
\end{figure}

In this subsection, we focus in particular on the case $\Sigma^2<0$ since, as we show later, the results in this case are very different from those when $\Sigma^2>0$. We look at the transmissivity in Eq.~\eqref{eq: transm dec equals}, where $\tau$ grows very slowly once $t\ge d$. Indeed, if we Taylor expand the expression inside the brackets in the right hand side of Eq.~\eqref{eq: transm dec equals} with respect to small $t-d$, we find that the lowest order term would be proportional to $(t-d)^4$. When this occurs, the transmissivity $\tau$ reaches a peak before decaying exponentially as $e^{-\gamma(t-d)}$. A plot of the transmissivity $\tau$ as function of time $t$ is presented in Fig.~\ref{tau1}. 

 Unfortunately, it is not possible to find an exact solution for the maximum of $\tau$ as a function of time. However, if $|\Sigma^2|\gg\gamma^2$ ($\gamma_\Sigma^2$ is negative in this case), the first term inside the bracket in Eq.~\eqref{eq: transm dec equals} becomes negligible. In this case, at times $(t-d)^2\gg|\gamma_\Sigma^2|$, the transmissivity \eqref{eq: transm dec equals} can be approximated as
\begin{align}
    \tau\sim-\frac{\gamma^2e^{-2\gamma (t-d)}}{d^2\gamma^2_\Sigma}(t-d)^2\theta(t-d)\,.
\end{align}
This equation allows us to find an approximate expression for the maximum of $\tau$. In particular, we find that $\tau$ reaches its maximum at the time $t_{max}\sim1/\gamma$ (which can also be seen in Fig.~\ref{tau1}). The maximum value reached by $\tau$ at time $t_{max}$ is
\begin{equation}\label{taumaximum}
    \tau(t_{max})\sim\frac{e^{-2}}{d^2\left(\omega^2-\sqrt{\frac{\pi}{8}}\frac{\gamma}{\sigma}-\gamma^2\right)}\,.
\end{equation}
From Eq.~\eqref{taumaximum} we can see how the peak of the transmissivity in time $\tau(t_{max})$ increases by increasing $\gamma$ keeping $\omega$ constant, or by decreasing $\omega$ keeping $\gamma$ constant.

\begin{figure}[ht!]
    \centering
    \includegraphics[width=0.9\linewidth]{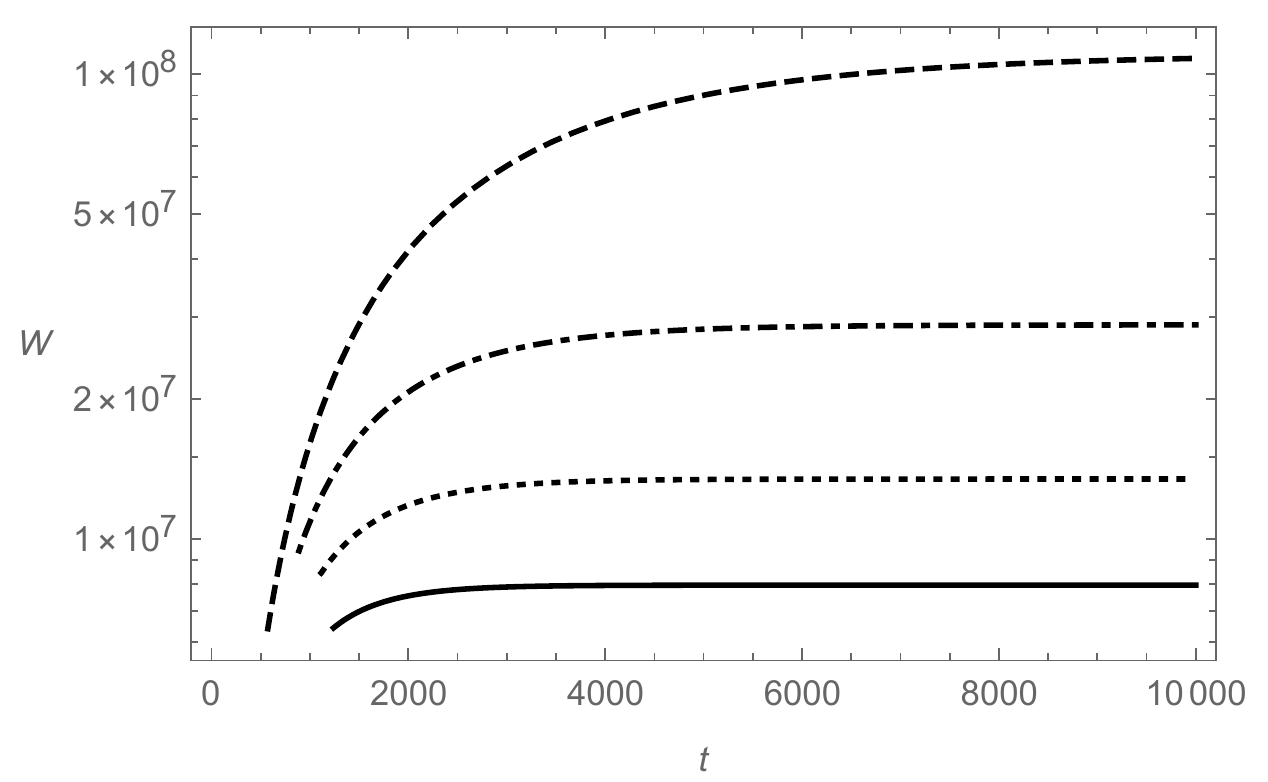}
    \caption{Determinant of the matrix $\mathbb{N}$, also called $W$, quantifying the noise at Bob's detector, as function of time for different values of $\gamma$ and when the parameter $\Sigma^2$ is negative. In particular $\gamma=1\cdot10^{-3}$ (thick line), $\gamma=0.75\cdot10^{-3}$ (dotted line), $\gamma=0.5\cdot10^{-3}$ (dot-dashed line), $\gamma=0.25\cdot10^{-3}$ (dashed line). The other parameters used are $\sigma=0.01$, $\omega=1$, $d=4$.} 
    \label{noise1}
\end{figure}

The behaviour of the noise $W$ as a function of time $t$, quantified with the determinant of the matrix $\mathbb{N}$ and explicitly reported in Eq.~\eqref{detN}, is shown in Fig.~\ref{noise1}. 

From it, we can see that, after a certain time, $W$ becomes approximately constant. By studying the long time limit $t\to\infty$ of Eq.~\eqref{detN}, an asymptotic expression can be analytically found, and it reads
\begin{equation}
    \lim_{t\to\infty}W=\frac{(4\gamma^2+|\Sigma^2|+2d^2\Sigma^4)(1+2d^2|\Sigma^2|)}{32\cdot64 \pi\sigma^2 d^4 \gamma^2\Sigma^8}\,.
\end{equation}
One can easily check that the asymptotic value of the noise decreases by increasing both $\gamma$ and $\omega$. Moreover, Fig.~\ref{noise1} informs us that the asymptotic value is reached at shorter times for larger $\gamma$.

\begin{figure}[ht!]
    \centering
    \includegraphics[width=0.9\linewidth]{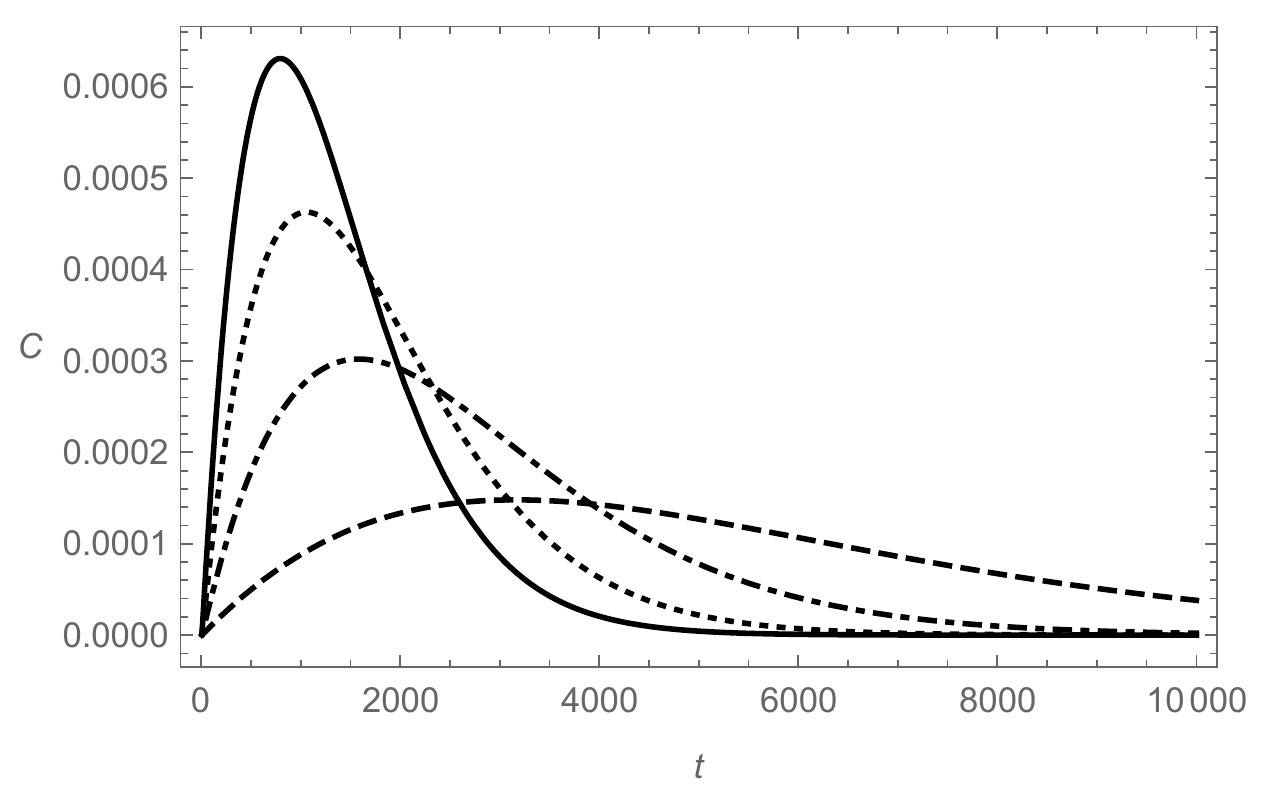}
    \caption{Lower bound of the classical capacity $C$ vs time when the parameter $\Sigma^2$ is negative. In particular $\gamma=1\cdot10^{-3}$ (thick line), $\gamma=0.75\cdot10^{-3}$ (dotted line), $\gamma=0.5\cdot10^{-3}$ (dot-dashed line), $\gamma=0.25\cdot10^{-3}$ (dashed line). The other parameters used are $\sigma=0.01$, $\omega=1$, $d=4$.}
    \label{capacity1}
\end{figure}

The lower bound for the classical capacity $C$, evaluated through Eq.~\eqref{CapacitySimplified}, is plotted in Fig.~\ref{capacity1} with encoding energy\footnote{The encoding energy $E$ was chosen to be $1/\sigma$ to compensate the smallness of the detectors. However, we remind the reader that $E$ can in principle be arbitrarily high, though finite.} $E=100$. 

\begin{figure}[ht!]
    \centering
    \includegraphics[width=0.9\linewidth]{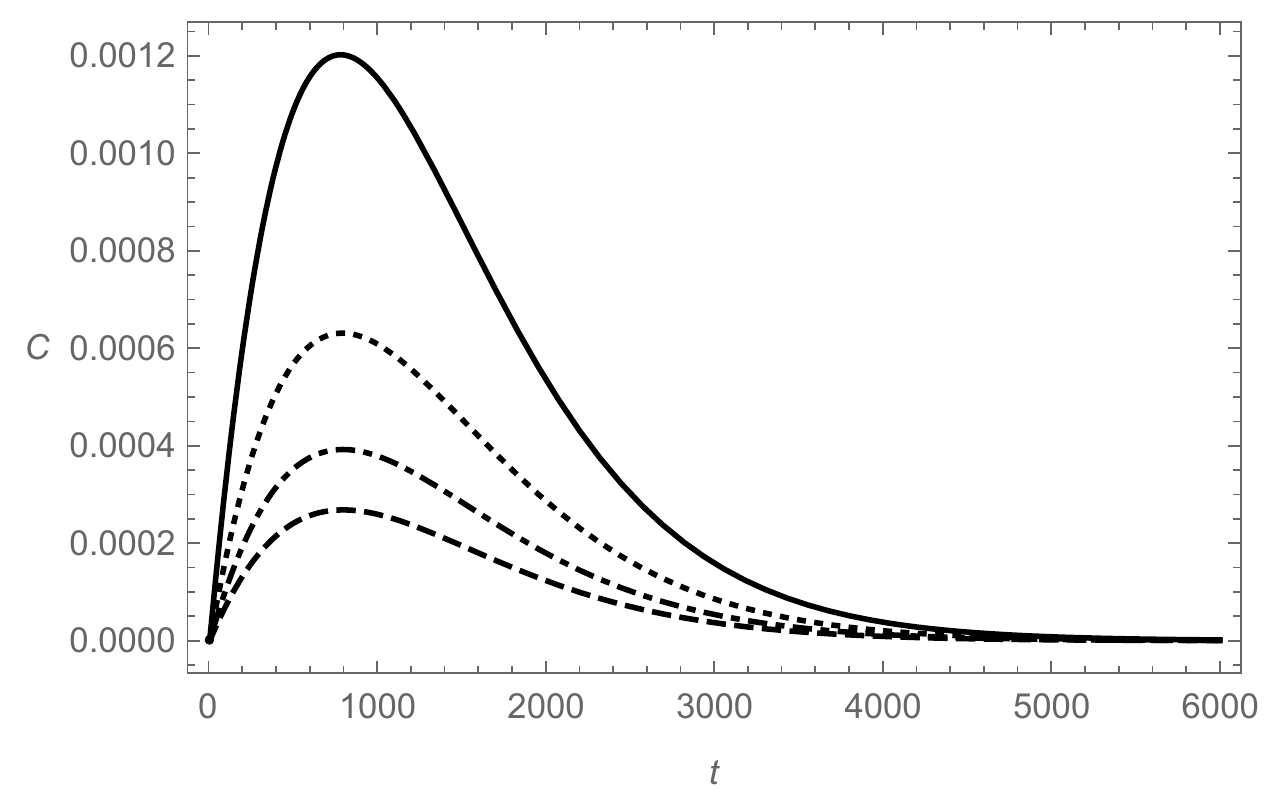}
    \caption{Lower bound of the classical capacity $C$ vs time when the parameter $\Sigma^2$ is negative. Different values for the detectors' frequency $\omega$ are used. In particular $\omega=0.75$ (thick line), $\omega=1$ (dotted line), $\omega=1.25$ (dot-dashed line), $\omega=1.5$ (dashed line). The other parameters used are $\sigma=0.01$, $\omega=1$ and $d=4$.}
    \label{Capacity2}
\end{figure}

The behaviour of $C$ in time is very similar to the one of $\tau$ shown in Fig.~\ref{tau1}. However, the maximum of $C$ is anticipated with respect to the one of $\tau$. This is because of the increasing of the noise occurring in time (the plateau the noise reaches comes later than the maximum of $\tau$). This effect is more pronounced for low values of $\gamma$. The classical capacity $C$ can be optimized for low values of $\omega$. The main reason for this is that, from the second equality of Eq.~\eqref{Energyconstraint}, the lower the magnitude of $\omega$, the higher the number of particles we can use for the encoding. We have plotted $C$ in Fig.~\ref{Capacity2}.

Summarizing, to optimize the communication capabilities of this protocol in the case  $-\Sigma^2\gg2\gamma/d$, increasing the frequency of the detectors is always inconvenient. Instead, increasing the value of the field-detector couplings $\gamma$ increases the maximum value of the capacity. However, we cannot increase $\gamma$ arbitrarily, since the condition $|\Sigma^2|\gg2\gamma/d$ must remain satisfied.

It is worth noticing from Fig.~\ref{capacity1} that for low values of $\gamma$ the capacity vanishes slower than in the case of high values of $\gamma$. In other words, even if the peak is wider for larger couplings $\gamma$, it is also more narrow. Therefore, in some particular situations (e.g., if we want the communication to last for a long time) it may be convenient to choose a lower value of $\gamma$.

Regarding the quantum capacity, as we mentioned in Sec.~\ref{quantumC}, a necessary condition for it to be different than zero is that $\tau>1/2$. In the case considered here, this condition is never reached and thus we conclude that the quantum capacity always vanishes.

\subsubsection{Case II: $|\Sigma^2|\gg2\gamma/d$ with $\Sigma^2>0$}
\begin{figure}[ht!]
    \centering
    \includegraphics[width=0.9\linewidth]{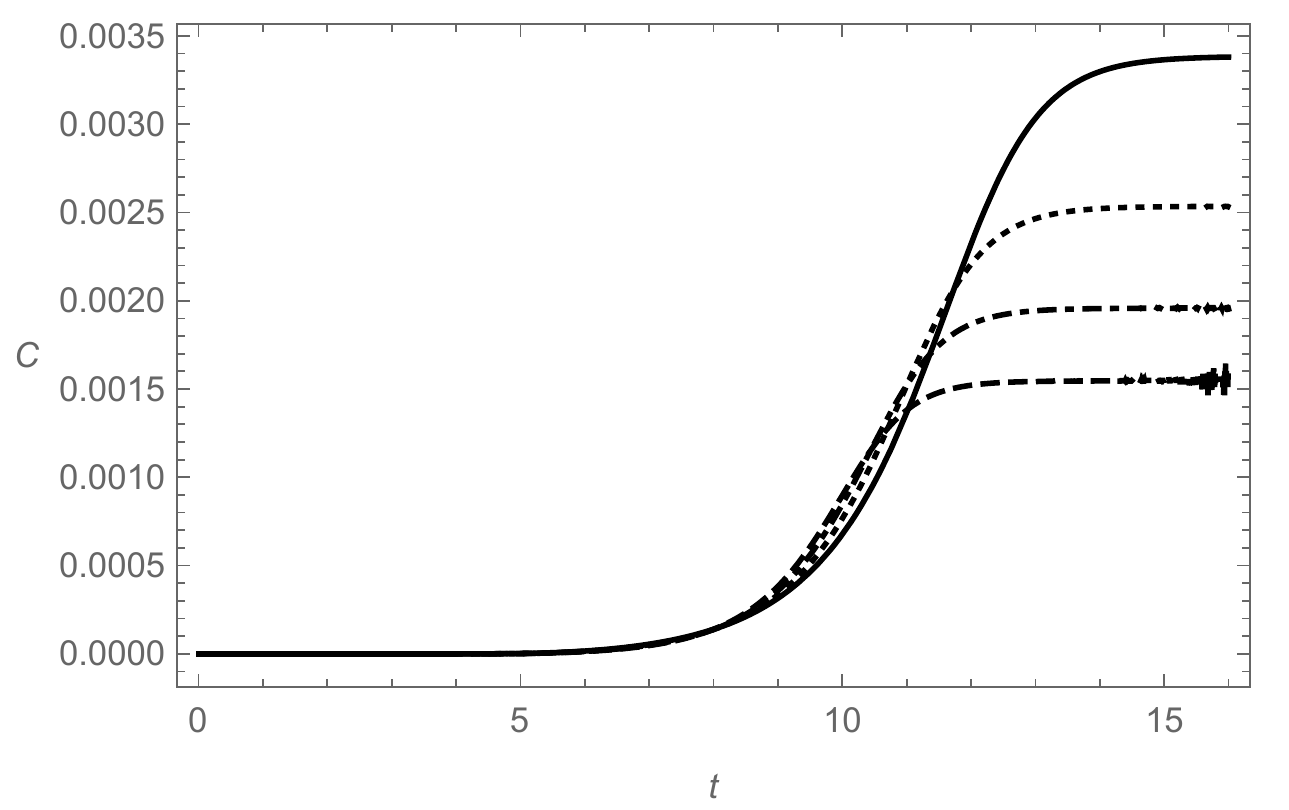}
    \caption{Lower bound of the classical capacity $C$ as function of time when the parameter $\Sigma^2$, defined after Eq.~\eqref{systemLangevin}, is positive. For the capacity $C$ Eq.~\eqref{CapacitySimplified} is used. Different values for the detectors' coupling with the field $\gamma$ are used. In particular, $\gamma=0.01$ (thick line), $\gamma=0.011$ (dotted line), $\gamma=0.012$ (dot-dashed line) and $\gamma=0.013$ (dashed line).}
    \label{capacity3}
\end{figure}

\begin{figure}[ht!]
    \centering
    \includegraphics[width=0.9\linewidth]{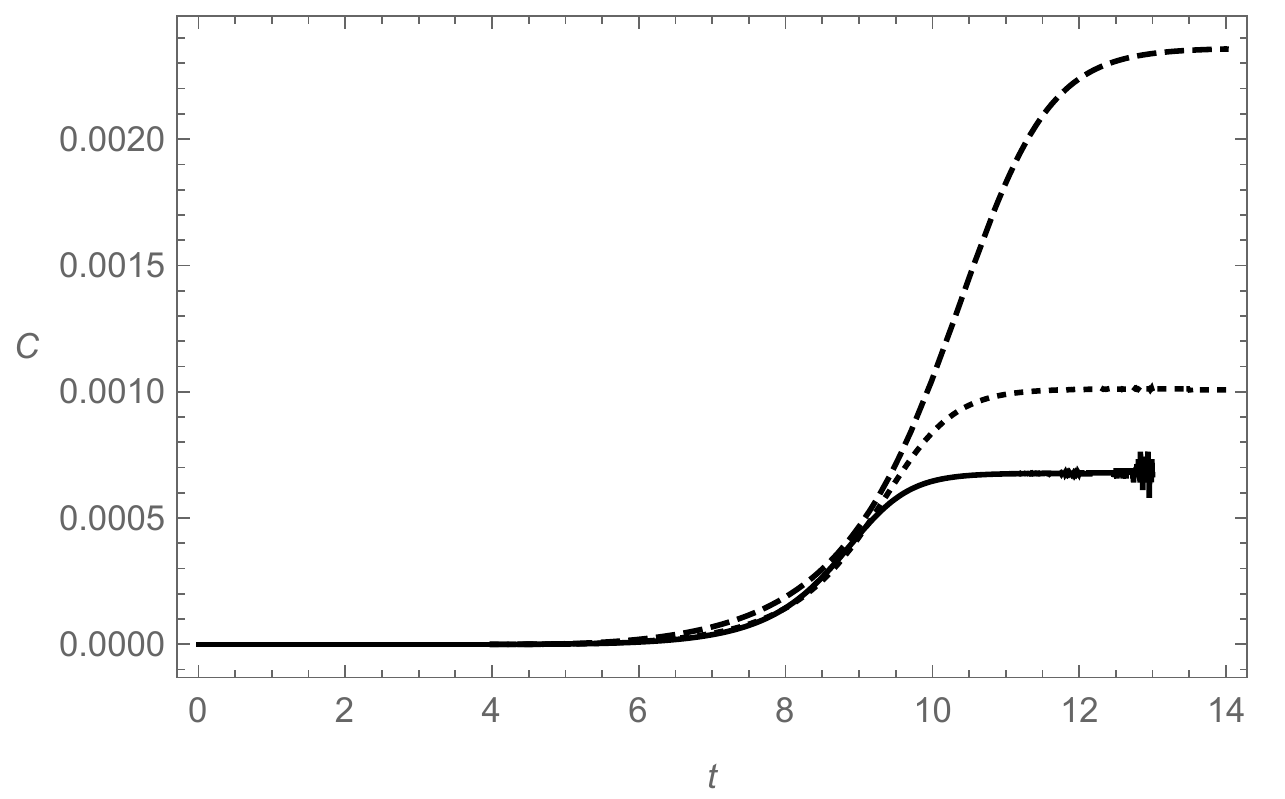}
    \caption{Lower bound of the classical capacity $C$ as function of time when the parameter $\Sigma^2$, defined after Eq.~\eqref{systemLangevin}, is positive. For the capacity $C$ Eq.~\eqref{CapacitySimplified} is used. Different values for the detectors' frequency $\omega$ are used. In particular, $\omega=0.75$ (thick line), $\omega=1$ (dotted line) and $\omega=1.25$ (dashed line).}
    \label{capacity4}
\end{figure}

We proceed to analyze the regime where $|\Sigma^2|\gg2\gamma/d$ with $\Sigma^2>0$. In this case, Eqs.~\eqref{eq: transm dec equals} and \eqref{detN} for the the transmissivity $\tau$ and for the noise $W$ respectively apply. Their dependence on the parameters is very different to the corresponding quantities for $\Sigma^2<0$. In fact, in this case, both $\tau$ and $W$ exponentially increase in time. At late times, we have
\begin{equation}
    \tau\sim\frac{\gamma^2}{4d^2\gamma_\Sigma^4}e^{2(t-d)\left(\gamma_\Sigma-\gamma\right)}\,,
\end{equation}
while, for the noise $W$ we have
\begin{align}
    W\sim&\gamma ^4 \left(\Sigma ^4 \gamma_\Sigma+8 \gamma ^4 \gamma_\Sigma+12 \gamma ^3 \Sigma^2 +8 \gamma ^2 \Sigma^2  \gamma_\Sigma+8 \gamma ^5\right.\nonumber\\
    &\left.+4 \gamma  \Sigma^4 +\frac{4 d^2 \Sigma ^4 \gamma_\Sigma^3 \left(\gamma+\gamma_\Sigma \right)^2 e^{2 d \left(\gamma_{\Sigma}-\gamma \right)}}{\gamma ^2}\right) \nonumber\\
    &\times \frac{e^{4 \left(\gamma_\Sigma-\gamma \right) (t-d)}}{256 d^4 \Sigma
   ^8 \gamma_\Sigma^9}\,.
\end{align}
Despite the exponential growth of $\tau$ and $W$ as a function of time, the capacity  asymptotically reaches a finite value as can be seen in Fig.~\ref{capacity3} and Fig.~\ref{capacity4}. 

From now on, we define $C_{\infty}\coloneqq C(t\to\infty)$.
An approximated expression for $C_{\infty}$ can be found. Indeed, Eq.~\eqref{CapacitySimplified} can be simplified using the fact that the function $h(x)\sim\log(x)$ when $x\gg1$. Since the noise $W$ is proved to be always very high, we can exploit its asymptotic behaviour. Therefore, further algebraic manipulations of Eq.~\eqref{CapacitySimplified} give us
\begin{equation}\label{classical capacity simplified}
    C\sim\log\left(1+\frac{\tau}{\sqrt{W}}\frac{E}{\omega}\right)-\log\left(1+\frac{\tau}{\sqrt{W}}\right)\,.
\end{equation}
Since $E/\omega$ is strictly positive, the capacity $C$ is a monotonic function of the ratio $\tau/\sqrt{W}$. The latter is finite for $t\to\infty$, and it has the expression \eqref{Asymptotic:capacity:appendix} that we do not report here for the sake of readability.
By inserting Eq.~\eqref{Asymptotic:capacity:appendix} into Eq.~\eqref{classical capacity simplified}, we get the value of the late time classical capacity $C_{\infty}$. This value increases by decreasing $\gamma$ and increasing $\omega$, optimizing the capacity. However, we cannot increase $\omega$ or decrease $\gamma$ arbitrarily since the condition $\Sigma^2=\sqrt{\frac{8}{\pi}}\frac{\gamma}{\sigma}-\omega^2\gg2\frac{\gamma}{d}$ must remain satisfied.

Finally, we use Eq.~\eqref{maxcoherentinf2} to estimate the quantum capacity. In the case considered here, since $\tau$ increases exponentially the first term vanishes at late times. Since the entropy function $h(\cdot)$ is strictly positive, $I_\textrm{c}$ becomes negative at late times which in turn implies that $Q^{(1)}$ vanishes as well. When $\tau\sim1$, the maximized coherent information assumes the form given by Eq.~\eqref{BmaximizedcoherentInf}. However, one can easily check that $\sqrt{W}\gg1$ when $\tau\sim1$, making $I_\textrm{c}$ negative and thus again $Q^{(1)}=0$.

\subsubsection{Case III: $|\Sigma^2|\approx2\gamma/d$}\label{resonancecase}
In this case we assume that $|\Sigma^2|$ is of the order of $2\gamma/d$. No approximations could be performed to solve Eq.~\eqref{systemLangevin}, even for identical detectors. All quantities have to be computed numerically (see Appendix~\ref{appendixA}). 

\begin{figure}[ht!]
    \centering
    \includegraphics[width=0.9\linewidth]{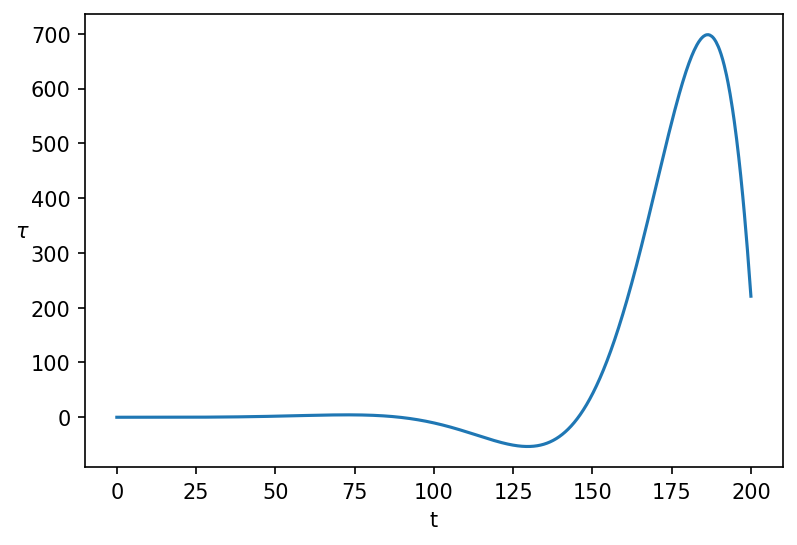}
    \caption{Plot of the transmissivity $\tau$ as function of time. The parameters used are $\sigma=0.01$, $\omega=1$, $d=4$ and $\gamma=\frac{\pi}{8}\sigma\omega^2$, so that $\Sigma^2$, defined after Eq.~\eqref{systemLangevin}, is null.}
    \label{tau2}
\end{figure}

We ask first what happens when $\Sigma^2=0$, since both the cases studied until now suggests that the capacity increases the smaller $|\Sigma^2|$ is. The behaviour of the transmissivity $\tau$, is depicted in Fig.~\ref{tau2}.

\begin{figure}[ht!]
    \centering
    \includegraphics[width=0.9\linewidth]{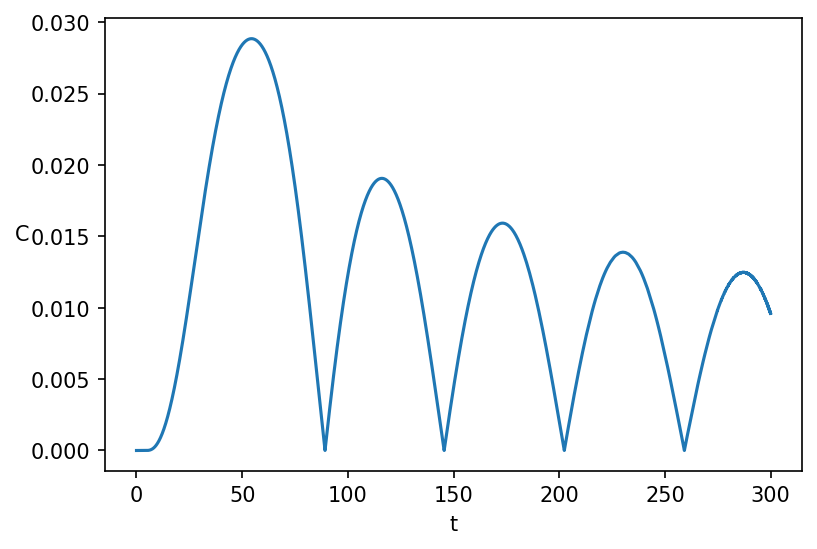}
    \caption{Lower bound of the classical capacity as function of time when $\Sigma^2$, defined after Eq.~\eqref{systemLangevin}, is null. The parameters used are $\sigma=0.01$, $\omega=1$, $d=4$ and $\gamma=\frac{\pi}{8}\sigma\omega^2$.}
    \label{capacityOsc}
\end{figure}

We see that $\tau$ oscillates between positive and negative values. The amplitude of these oscillations grows exponentially with the time. When $\tau$ is negative, the quantum channel behaves as the conjugate of a linear amplifier. Nevertheless, the classical capacity is not affected by the sign of $\tau$ when $r=1$. Indeed, it has been shown that, when $\tau$ is negative, the expression \eqref{CapacitySimplified} holds replacing $\tau$ with $|\tau|$, see \cite{Lupo_2011}.
Fig.~\ref{capacityOsc} shows the behaviour of the classical capacity when $\Sigma^2=0$. 

As a consequence of the oscillation of $\tau$, the capacity has corresponding peaks with all positive values. This time, the amplitudes of the peaks decrease exponentially with $t$. Thus, for $t\to\infty$ we have $C\to0$. 

\begin{figure}
    \centering
    \includegraphics[width=0.9\linewidth]{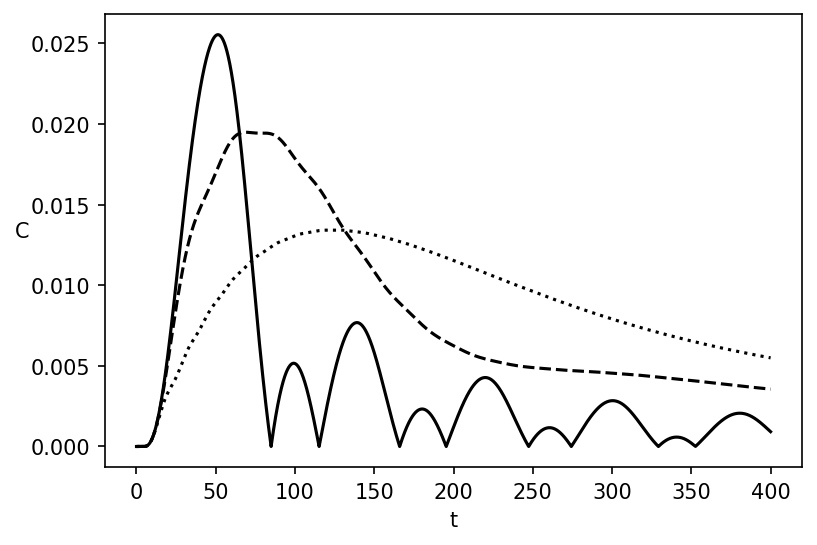}
    \caption{Lower bound of the classical capacity as function of time for small negative values of $\epsilon$, defined in Eq.~\eqref{epsilon} and encoding energy $100$. In particular $\epsilon=-10^{-3.5}$ (dotted line), $\epsilon=-10^{-4.1}$ (dashed line) and $\epsilon=-10^{-4.7}$ (solid line). The parameters used are $\sigma=0.01$, $\omega=1$, $d=4$ and $\gamma=\frac{\pi}{8}\sigma\omega^2$.}
    \label{capacity5}
\end{figure}

\begin{figure}
    \centering
    \includegraphics[width=0.9\linewidth]{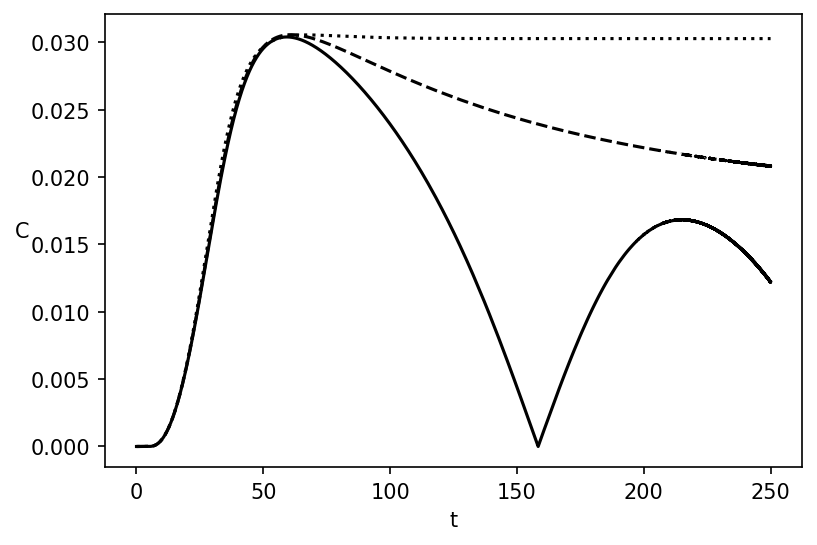}
    \caption{Lower bound of the classical capacity as function of time for small positive values of $\epsilon$, defined in Eq.~\eqref{epsilon} and encoding energy $100$. In particular $\epsilon=10^{-4.5}$ (dotted line), $\epsilon=10^{-4.7}$ (dashed line) and $\epsilon=10^{-4.8}$ (solid line). The parameters used are $\sigma=0.01$, $\omega=1$, $d=4$ and $\gamma=\frac{\pi}{8}\sigma\omega^2$.}
    \label{capacity6}
\end{figure}

We proceed next to study what happens when $\Sigma^2$ is slightly negative or slightly positive. Our capacity $C$ is now plotted in Figs. \ref{capacity5} and \ref{capacity6}, respectively for such cases.

To plot the capacity we have chosen the coupling $\gamma$ of the form
\begin{equation}\label{epsilon}
    \gamma=\sqrt{\frac{\pi}{8}}\sigma\omega^2+\epsilon\,,
\end{equation}
for convenience, where $\epsilon$ is a free parameter that we will vary. The value of $\Sigma^2$ is therefore $\Sigma^2=\sqrt{\frac{8}{\pi}}\frac{\epsilon}{\sigma}$.

Comparing the capacities in Figs. \ref{capacity1} and \ref{capacity5}, we can see how the transition between the behaviour when $|\Sigma^2|\gg\frac{\gamma}{d}$ and the one when $|\Sigma^2|\sim\frac{\gamma}{d}$ occurs for very small values of $\Sigma^2$. In other words, for the parameters chosen, even when $\epsilon=-10^{-3.5}$ ($|\Sigma^2|\sim 32\cdot\gamma/d$) the capacity qualitatively behaves as when $|\Sigma^2|\gg\gamma/d$ (shown in Fig.~\ref{capacity1}). When $\epsilon=-10^{-4.1}$ ($|\Sigma^2|\sim8$), a small oscillatory behaviour appears in the capacity as shown in Fig.~\ref{capacity5}. For $\epsilon=-10^{-4.7}$ the oscillating behaviour is fully present. It is worth specifying that, even in this case, the capacity decreases to zero with increasing time. We conclude that this is a general property valid for $\Sigma^2\le0$. 

For $\Sigma^2\sim2\gamma/d$ and positive, we have the capacities in Fig.~\ref{capacity6}. Again, the capacity qualitatively behaves as the case $\Sigma^2\gg2\gamma/d$ even when $\epsilon=10^{-4.5}$ ($\Sigma^2\sim3\gamma/d$). Then the capacity starts to decrease after the peak, until it reaches a negative value and starts to oscillate. In this case, for $t\to\infty$ a finite value of the capacity is expected. However, a numerical study of this value is inhibited by the enormous chattering occurring at late times.

\begin{figure}[ht!]
    \centering
    \includegraphics[scale=1]{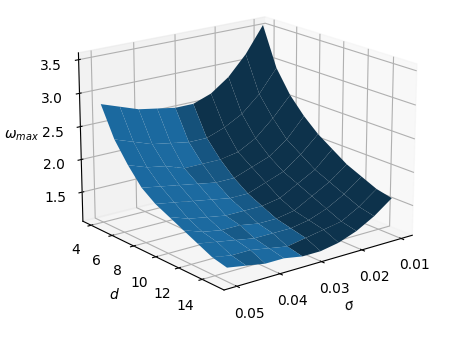}
    \caption{Plot of the frequency $\omega_{max}$ needed to optimize the late time classical capacity $C_{\infty}$, following the criterion explained under Eq.~\eqref{relationBestSetup}, for different values of $\sigma$ and $d$. The coupling is $\gamma=\gamma_{max}(\omega_{max})$, following Eq.~\eqref{relationBestSetup}.}
    \label{maximizingFrequency}
\end{figure}

The ideal situation is the one in which the classical capacity remains constant after reaching its maximum, becoming $C_\infty$. Therefore, the best situation shown in Fig.~\ref{capacity6} is the capacity behaviour of the dotted line, corresponding to $\Sigma^2\sim3\gamma/d$. By testing different values of $\Sigma^2$ even for different setups (different $\sigma$ or $d$), the best value of $C_{\infty}$ occurs always when $\Sigma^2$ is very close to $4\gamma/d$. In particular, testing the values for $d$ and $\sigma$ used in Fig.~\ref{maximizingFrequency}, the standard deviation of $\frac{d\Sigma^2}{\gamma}$ from the value $4$ is of order $10^{-4}$, which can be considered a negligible numerical error. We can therefore conclude that, in order to reach the asymptotic constant value $C_{\infty}$, we must require $\Sigma^2=4\gamma/d$. In term of the parameters of the problem this condition reads 
\begin{equation}\label{relationBestSetup}
    \gamma=\gamma_{max}(\omega)\coloneqq\frac{d}{4}\frac{\omega^2}{\left(\sqrt{\frac{8}{\pi}}\frac{d}{4\sigma}-1\right)}\,.
\end{equation}

Summarizing, once we fix $\omega$, $d$ and $\sigma$, if we choose $\gamma=\gamma_{max}(\omega)$ then the classical capacity $C$ increases in time reaching an asymptotic value $C_{\infty}$. If $\gamma<\gamma_{max}$, the capacity $C$ starts to decrease after reaching its maximum value $\sim C_{\infty}$, as shown in Fig.~\ref{capacity6}, therefore having an asymptotic value lower than $C_{\infty}$. If $\gamma>\gamma_{max}$, the growth of $C$ stops earlier with respect to the case $\gamma=\gamma_{max}$ reaching, again, an asymptotic value lower than $C_\infty$. 

\begin{figure}[ht!]
    \centering
    \includegraphics[width=0.9\linewidth]{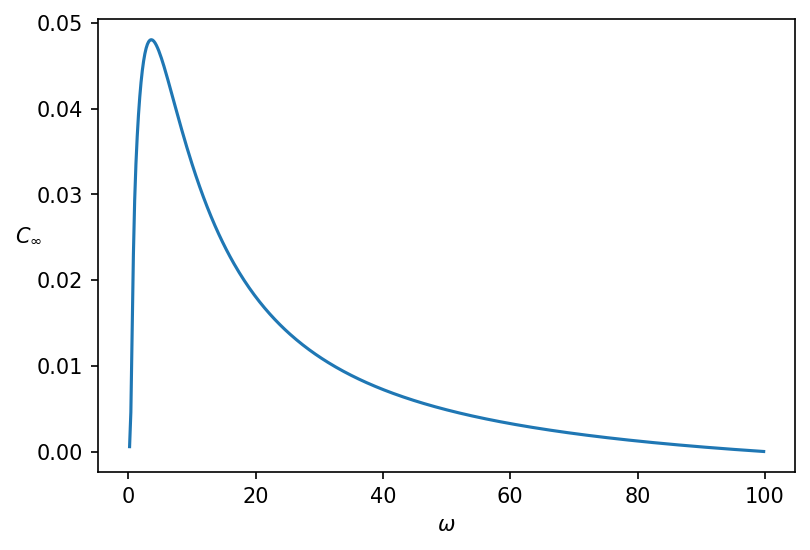}
    \caption{Asymptotic value of the capacity $C$ (also called $C_{\infty}$) as a function of $\omega$, when $\gamma=\gamma_{max}(\omega)$. The other parameters chosen are $d=4$ and $\sigma=0.01$.}
    \label{maxCapacity}
\end{figure}

We now ask what frequency $\omega$ should we choose in order to have the asymptotic value $C_{\infty}$ as large as possible. One can numerically see that the ratio between $\tau$ and $\sqrt{W}$ increases with growing the detectors' frequency $\omega$. However, if we fix the encoding energy $E$, the larger the frequency $\omega_\textrm{A}=\omega$ of Alice's detector, the more encoding energy would be spent to prepare Alice's initial state (this can be explicitly seen in Eq.~\eqref{CapacitySimplified}). As a consequence, it is shown in Fig.~\ref{maxCapacity} that it is convenient to increase the detectors' frequency $\omega$, as well as the coupling $\gamma=\gamma_{max}(\omega)$, only until a certain value, which we call $\omega_{max}$. After this value, increasing the detectors' frequency becomes inconvenient, because the loss we would have decreasing the number of encoding particles $N=E/\omega_\textrm{A}-1/2$ overcomes than the gain obtained increasing $\tau/\sqrt{W}$.

With the parameters $\sigma$ and $d$ chosen in Fig.~\ref{maxCapacity}, the detector frequency maximizing  $C_{\infty}$ (which we call $\omega_{max}$ from now on) is $\omega_{max}\simeq3.4$. 
The value $\omega_{max}$ is numerically found also for other configurations of $\sigma$ and $d$. The result for $\omega_{max}$ is shown in Fig.~\ref{maximizingFrequency}.
Hence, for the given values of $\sigma$ and $d$ in Fig.~\ref{maximizingFrequency}, we found the values of the detector frequency and coupling necessary to have the best classical capacity. It is possible to repeat the numerical procedure with every protocol setup (i.e., for every value of $\sigma$ and $d$). In this way, we always know how to tune the detectors' frequency and the coupling in order to maximize the communication of classical messages.

Finally, we numerically computed the maximized coherent information $I_\textrm{c}$ from Eq.~\eqref{maxcoherentinf2}. Even in the best classical capacity scenario, it results always negative. We think that this occurs because, in all the contexts considered, we obtained $\sqrt{W}\gg\tau$ thus making the second term of $I_\textrm{c}$ in Eq.~\eqref{maxcoherentinf2} much larger than the first. As a consequence, the quantum capacity turns out to be always zero for identical detectors. We leave it to the future  to investigate the possibility of a reliable communication of quantum messages by dropping the approximation $\sigma\ll d$.

\subsection{Different detectors}\label{subsec: Different detectors}
We now proceed with the study of detectors that have different frequencies and/or couplings with the field, namely $\omega_\textrm{A}\ne\omega_B$ and $\gamma_\textrm{A}\ne\gamma_\textrm{B}$. In Sec.~\ref{resonancecase} we have seen that the capacity for identical detectors is optimized when $|\Sigma^2|$ is comparable to $2\gamma/d$, therefore we make an educated guess and here limit ourselves to setups where $|\Sigma_\textrm{A}^2|\sim2\gamma_\textrm{A}/d$ and $|\Sigma_\textrm{B}^2|\sim2\gamma_\textrm{B}/d$, leaving the result for $\tau$ for other setups in appendix \ref{appendixD}. The main motivation is that our goal is to find a setup maximizing the classical capacity. Henceforth, we want to see how the best classical capacity obtained in the identical detector case changes by making the detectors with different parameters $\omega_i$ and $\gamma_i$.

\subsubsection{Case I: $\gamma_\textrm{A}\ne\gamma_\textrm{B}$ with $\omega_\textrm{A}=\omega_B$}
\begin{figure}[ht!]
    \centering
    \includegraphics[width=0.9\linewidth]{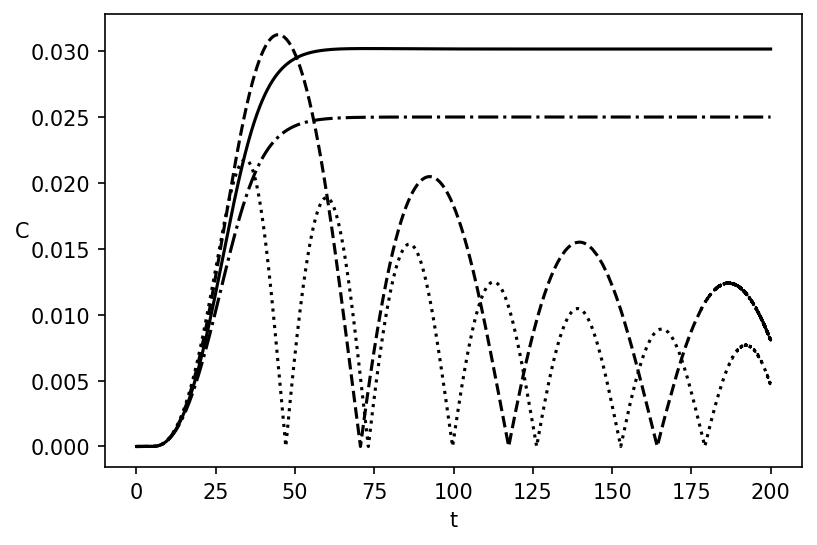}
    \caption{Lower bound of the classical capacity with encoding energy $E=100$, as function of time $t$, for values $\gamma_\textrm{B}$ different than $\gamma_\textrm{A}$. In particular, $\gamma_\textrm{B}=1.01\cdot\gamma_\textrm{A}$ (dot-dashed line), $\gamma_\textrm{B}=\gamma_\textrm{A}$ (solid line), $\gamma_\textrm{B}=0.99\cdot\gamma_\textrm{A}$ (dashed line) and $\gamma_\textrm{B}=0.98\cdot\gamma_\textrm{A}$ (dotted line). The other parameters chosen are $d=4$, $\sigma=0.01$, $\omega_\textrm{A}=\omega_B=1$ and $\gamma_\textrm{A}=\gamma_{max}(\omega_\textrm{A})$, where $\gamma_{max}$ is defined in Eq. \eqref{relationBestSetup}.}
    \label{CapacityDD1}
\end{figure}

We first consider the case with different detector-field couplings $\gamma_\textrm{A}\ne\gamma_\textrm{B}$, but with equal frequencies $\omega=\omega_\textrm{A}=\omega_B$. In  Fig.~\ref{CapacityDD1}, we present the lower bound to the classical capacity $C$, using $\gamma_\textrm{A}=\gamma_{max}(\omega_\textrm{A})$ and different values of $\gamma_\textrm{B}$ in the neighborhood of $\gamma_{max}(\omega_B)$. When $\gamma_\textrm{B}>\gamma_\textrm{A}$, we can observe that the late time capacity $C_\infty$ is lower than the case $\gamma_\textrm{A}=\gamma_\textrm{B}$. 

Instead, when $\gamma_\textrm{B}<\gamma_\textrm{A}$, the capacity presents an oscillating behaviour, which means that it cannot always be nonzero at late times. The frequency of the capacity oscillations is proportional to $\gamma_\textrm{A}-\gamma_\textrm{B}$. Furthermore, comparing the dashed and the solid lines, we observe that, when $\gamma_\textrm{B}$ is slightly smaller than $\gamma_\textrm{A}$, there is a period of time in which the capacity is higher with respect to the identical detectors case. Nevertheless, the capacity drops to zero at late times and, for this reason, the setup $\gamma_\textrm{A}=\gamma_\textrm{B}$ may be the preferable one for communication.

\subsubsection{Case II: $\gamma_\textrm{A}\ne\gamma_\textrm{B}$ and $\omega_\textrm{A}\ne\omega_B$}
We now consider the case of different detector frequencies, and define $\omega_\textrm{A}:=\overline{\omega}/\alpha$ and $\omega_B:=\alpha \overline{\omega}$ for convenience. Moreover, we assume that the couplings are the optimal ones, i.e., $\gamma_\textrm{A}=\gamma_{max}(\omega_\textrm{A})$ and $\gamma_{B}=\gamma_{max}(\omega_B)$.

\begin{figure}[ht!]
    \centering
    \includegraphics[width=0.9\linewidth]{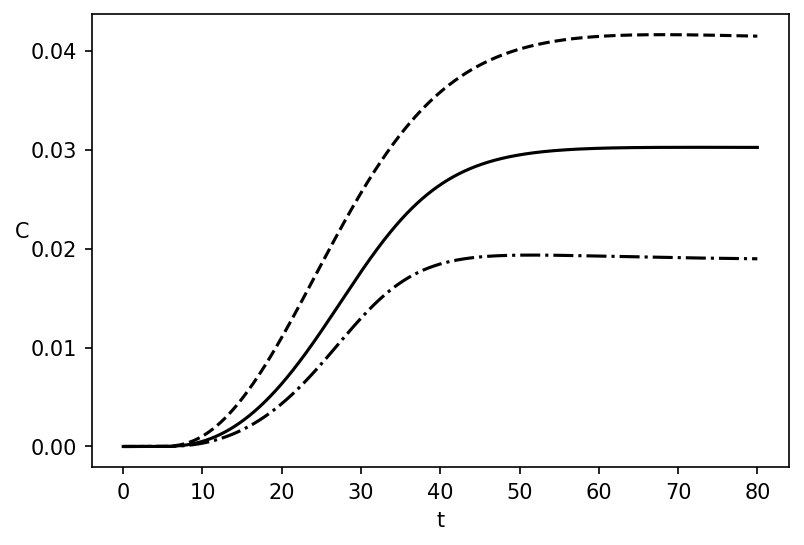}
    \caption{Lower bound of the classical capacity with fixed encoding energy $E=100$, as function of the time $t$, for different values of the parameter $\alpha$, changing the frequencies and the couplings of the detectors. In particular, $\alpha=1.5$ (dashed line), $\alpha=1$ (solid line), $\alpha=0.5$ (dot-dashed line). The other parameters are $\overline{\omega}=1$, $\sigma=0.01$, $d=4$.}
    \label{CapacityDD2}
\end{figure}

In this case,  numerical results for the capacity are shown in Fig.~\ref{CapacityDD2}. 

\begin{figure}[ht!]
    \centering
    \includegraphics[width=0.9\linewidth]{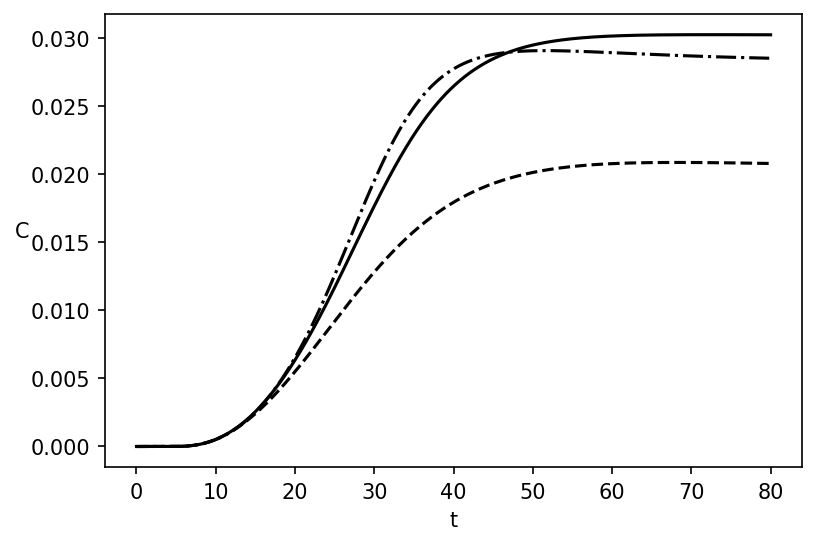}
    \caption{Lower bound of the classical capacity with fixed encoding number of particles $N=100$, as function of the time $t$, for different values of the parameter $\alpha$, changing the frequencies and the couplings of the detectors. In particular, $\alpha=1.5$ (dashed line), $\alpha=1$ (solid line), $\alpha=0.5$ (dot-dashed line). The other parameters are $\overline{\omega}=1$, $\sigma=0.01$, $d=4$.}
    \label{CapacityDD3}
\end{figure}

We can see that the capacity increases when $\omega_B>\omega_\textrm{A}$. In fact, by decreasing Alice's detector frequency $\omega_\textrm{A}$, a greater number of encoding particles $N=E/\omega_\textrm{A}-1/2$ can be used for the communication protocol. In other words, more encoding energy is saved for this purpose. We now ask if the capacity increases for $\omega_B>\omega_\textrm{A}$ even if we keep the number of encoding particles $N$ fixed. Removing the explicit dependence of $\omega_\textrm{A}$ on the capacity, we have
\begin{equation}\label{CapacitySimplified2}
    C=h\left(\frac{\tau}{2}+\sqrt{W}+\tau N\right)-h\left(\frac{\tau}{2}+\sqrt{W}\right)\,.
\end{equation}
The result is shown in Fig.~\eqref{CapacityDD3}. 

This confirms that, by fixing the number of encoding particles, the identical detector setup remains the best for the communication protocol. It is worth noticing that, when $\omega_\textrm{A}>\omega_B$, the increase of the capacity is anticipated with respect to the case $\omega_\textrm{A}=\omega_B$. As a consequence, despite the asymptotic value of the capacity $C_{\infty}$ is lower compared to the one in the case of the identical detectors, the early time capacity is higher when $\omega_\textrm{A}>\omega_B$. Therefore, if we have constrained to $N$ the number of particles used, but we do not have a limit for the encoding energy $E$, it is convenient to set $\omega_\textrm{A}>\omega_B$ in situations in which one requires a good communication at early times. 

Finally, as the case in Sec.~\ref{resonancecase}, we never find a situation in which the maximized coherent information $I_\textrm{c}$, given by Eq.~\eqref{maxcoherentinf2}, is positive. This means that the quantum capacity $Q^{(1)}$ results zero also in this case.

\section{Conclusions}\label{conclusions}
In this work we studied the communication of classical and quantum messages between two field detectors, modelled as quantum oscillators, that are separated by a distance $d$, have characteristic sizes $\sigma$, and have frequencies $\omega_\textrm{A}$ (sender) and $\omega_B$ (receiver). The  communication channel is mediated by a scalar field that is coupled with both detectors via a monopole interaction, governed by the coupling constants $\gamma_\textrm{A}$ and $\gamma_\textrm{B}$ respectively. We focused on the communication of classical messages, quantified by the classical capacity of the communication channel, since we have shown that reliable communication of quantum messages is shown to be impossible for pointlike detectors (i.e., when $\sigma\ll d$).

In principle, one may expect  that the communication improves with increasing coupling between the detector and the field. In fact, we can say that a stronger coupling of the detectors with the field means the message to be communicated ``is better coupled'' with Alice and Bob's detector. However, the stronger Bob's coupling $\gamma_\textrm{B}$, the more Bob's detector witnesses noisy particles as well. To solve this problem, a strategy would be to decrease Bob's coupling $\gamma_\textrm{B}$, leaving a high coupling $\gamma_\textrm{A}$ in the case of Alice's detector. If the scalar field is coupled to a two-level Unruh-DeWitt detector, this strategy can be shown to work \cite{Landulfo2021,Tjoa2022}. However, we have shown that, for harmonic oscillator detectors, the communication properties are compromised even if we slightly deviate  from the equal coupling case $\gamma_\textrm{B}=\gamma_\textrm{A}$. As a consequence, the best setup in terms of magnitude of the capacity occurs when the detectors are identical and when their coupling $\gamma$ is equal to a finite value related to the other parameters $\sigma$, $d$ and $\omega$ through Eq.~\eqref{relationBestSetup}.

We have shown that the symmetric setup is the best in terms of maximizing the channel capacity. Nevertheless, it is worth mentioning another setup that makes the communication of a classical message faster at the expense of reliability. As shown in Fig.~\ref{CapacityDD3}, if we consider different detectors by changing both the frequencies and the field couplings, each one satisfying the constraint Eq.~\eqref{relationBestSetup}, the capacity exceeds that of the previous scenario at early times when $\gamma_\textrm{B}<\gamma_\textrm{A}$. This is valid exclusively if Alice has a limited amount of encoding particles but an unlimited amount of encoding energy.

The advantage of using oscillator-like detectors over qubit UDW detectors to communicate classical messages relies on the arbitrary (though finite) energy that can be used in the encoding process. In other words, if we have enough energy it is always possible to have a reliable communication of classical signals. For this reason it would be interesting to explore the advantage that oscillator-like detectors offer in the context of curved spacetime backgrounds or when they follow non-inertial trajectories \cite{Letaw}.

It may be worthwhile to investigate the feasibility of incorporating smooth time-dependent switching functions $\lambda (t)$ (see, e.g., \cite{Satz,Jorma:Satz}), which smoothly turn on and off the interaction between the detectors and the field, as a means to reduce the acquired noise. However, in this case, the coefficients $\gamma_i$ and $\Sigma_i^2$ become time-dependent and non-linear. Hence, the solution of the Langevin equation \eqref{systemLangevin} becomes challenging, even numerically. For this reason, we defer discussion of the potential to increase channel capacity through the use of smooth switching functions to future work.

The reliable communication of quantum signal seems to be impossible for the protocol studied. In future work, we aim to devise setups that maximize the coherent information and allow the possibility of a reliable communication of quantum messages. 

To conclude, we have quantified the classical channel capacity for communication protocols where a signal is communicated between two oscillator-like detectors by means of a scalar field. We believe that this is the first step in the direction of understanding (quantum) communication processes in relativistic contexts.

\section{Acknowledgments}
A. L. is grateful to Salvatore Capozziello and Orlando Luongo for helpful discussions. D. E. B. acknowledges support from the joint project No. 13N15685 ``German Quantum Computer based on Superconducting Qubits (GeQCoS)" sponsored by the German Federal Ministry of Education and Research (BMBF) under the framework ``Quantum technologies–from basic research to the market". S. M. acknowledges financial support from
``PNRR MUR project PE0000023-NQSTI".

\bibliographystyle{apsrev4-2}
\bibliography{references}

\newpage

\onecolumngrid

\appendix

\section{Dissipation and noise kernels for Gaussian smearing}\label{NewAppendixA}
In this appendix we report the expressions for Gaussian detectors without performing the point-like limit approximation \eqref{approximation}. The Gaussian spatial profile is
\begin{equation}\label{gaussian:shape:appendix}
    f_i(\xx-\xx_i)=\frac{1}{(\sqrt{\pi}\sigma)^3}e^{-(\xx-\xx_i)^2/\sigma^2}.
\end{equation}
We note that Gaussian functions do not have compact support and therefore the two detectors may directly interact with each other through the tails of the Gaussian. However,  by placing the detectors far enough each other, this direct cross-talk is suppressed exponentially thus assuring that at $t=0$ their quantum state can be taken to be uncorrelated, i.e., $\sigma_{\textrm{AB}}(0)=0$.

We calculate the elements of the dissipation kernel \eqref{disskernel} to obtain
\begin{align}\label{dissKernelRealoffD}
     \chi_{\textrm{AB}}(t)&=\chi_{\textrm{BA}}(t)=\frac{\lambda_\textrm{A}\lambda_\textrm{B}}{4\pi^2\sigma d}\theta(t)\sqrt{\frac{\pi}{2}}\left(e^{-\frac{(t-d)^2}{2\sigma^2}}-e^{-\frac{(t+d)^2}{2\sigma^2}}\right),
\end{align}
and the diagonal expression
\begin{equation}\label{dissKernelRealD}
    \chi_{KK}(t)=\frac{\lambda_K^2 t}{2\pi^2\sigma^3}\sqrt{\frac{\pi}{2}}e^{-\frac{t^2}{2\sigma^2}},
\end{equation}
where $d=|\mathbf{x}_A-\mathbf{x}_B|$ is the spatial distance between the two detectors, we have made the change of variable $t\to t-t'$, and $K=$A,B. 

The Fourier transform of the dissipation kernel elements are then given respectively by
\begin{align}
    \widetilde{\chi}_{\textrm{AB}}(z)&=\widetilde{\chi}_{\textrm{BA}}(z)=\frac{\sqrt{\gamma_\textrm{A}\gamma_\textrm{B}}}{d}e^{-\frac{-z^2\sigma^2}{2}}\left\{e^{{\rm i}z d}\left[1
    +\text{erf}\left(\frac{1+{\rm i}z\sigma}{\sqrt{2}}\right)\right]-e^{-{\rm i}z d}\left(1-\text{erf}\left(\frac{1-{\rm i}z\sigma}{\sqrt{2}}\right)\right)
    \right\},
\end{align}
and the diagonal term
\begin{align}\label{FTdisskernel}
     \widetilde{\chi}_{KK}(z)=\frac{\gamma_i}{\sigma}\sqrt{\frac{2}{\pi}}\bigg(2&+{\rm i}\sqrt{2\pi}z\sigma e^{-\frac{-z^2\sigma^2}{2}}-2\sqrt{2}z\sigma D\left(\frac{z \sigma}{\sqrt{2}}\right)\bigg),
\end{align}
where $\gamma_K:=\lambda_i^2/8\pi$, $\text{erf}(z)$ is the error function, and $D(z):=e^{-z^2}\int_{0}^z e^{t^2}dt$
is the Dawson's integral \cite{NIST}. On the other hand, the elements of the noise kernel \eqref{noisekernel} are 
\label{noisekernelRealoffD}
\begin{align}
    \nu_{\textrm{AB}}(t)&=\nu_{\textrm{BA}}(t)=\frac{\sqrt{2\gamma_\textrm{A}\gamma_\textrm{B}}}{\pi\sigma d}\left(D\left(\frac{t+d}{\sqrt{2}\sigma}\right)-D\left(\frac{t-d}{\sqrt{2}\sigma}\right)\right),
    \end{align}
and
\begin{equation}\label{noisekernelRealD}
    \nu_{KK}(t)=\frac{\gamma_K}{\pi\sigma^2}\left(1-\frac{\sqrt{2}t}{\sigma}D\left(\frac{t}{\sqrt{2}\sigma}\right)\right)
\end{equation}
in the case of the diagonal term.

\section{Green function calculation}\label{appendixA}
In this appendix, we clarify how to obtain the Green functions $G_{KK}(t)$, with K$=\textrm{A},\textrm{B}$. The set of equations to be solved is
\begin{align}\label{systemLangevin:appendix}
    \ddot{G}_{\textrm{AA}}(t)-\Sigma_\textrm{A}^2G_{\textrm{AA}}(t)+2\gamma_\textrm{A}\dot{G}_{\textrm{AA}}(t)=&2\frac{\sqrt{\gamma_\textrm{A}\gamma_\textrm{B}}}{d}G_{\textrm{AB}}(t-d)\theta(t-d)\,;\nonumber\\
    \ddot{G}_{\textrm{AB}}(t)-\Sigma_\textrm{A}^2G_{\textrm{AB}}(t)+2\gamma_\textrm{A}\dot{G}_{\textrm{AB}}(t)=&2\frac{\sqrt{\gamma_\textrm{A}\gamma_\textrm{B}}}{d}G_{\textrm{BB}}(t-d)\theta(t-d)\,;\nonumber\\
    \ddot{G}_{\textrm{BA}}(t)-\Sigma_\textrm{B}^2G_{\textrm{BA}}(t)+2\gamma_\textrm{B}\dot{G}_{\textrm{BA}}(t)=&2\frac{\sqrt{\gamma_\textrm{A}\gamma_\textrm{B}}}{d}G_{\textrm{AA}}(t-d)\theta(t-d)\,;\nonumber\\
    \ddot{G}_{\textrm{BB}}(t)-\Sigma_\textrm{B}^2G_{\textrm{BB}}(t)+2\gamma_\textrm{B}\dot{G}_{\textrm{BB}}(t)=&2\frac{\sqrt{\gamma_\textrm{A}\gamma_\textrm{B}}}{d}G_{\textrm{BA}}(t-d)\theta(t-d)\,,
\end{align}
from which it can be seen that $G_{\textrm{AB}}(t)=G_{\textrm{BA}}(t)$. The latter can be solved by taking step by step the range of times $0<t<d$, $d<t<2d$, $2d<t<3d$ etc. In the range of times $0<t<d$, the external forces are zero by the presence of the Heaviside theta in the right hand sides of Eqs.~\eqref{systemLangevin:appendix}. Therefore, all the elements of the Green function matrix behave as a free damped harmonic oscillator. 

For the diagonal elements $G_{KK}$, using the boundary conditions $\dot{G}_{KK}(0)=1$ and $G_{KK}(0)=0$, we obtain
\begin{equation}\label{GreenAA1st}
    G_{KK}(t)=\theta(t)\frac{e^{-\gamma_K t}\sinh(\sqrt{\gamma_K^2+\Sigma_K^2})t}{\sqrt{\gamma_K^2+\Sigma_K^2}}\,.
\end{equation}
At this point, we can calculate the Green function matrix elements $G_{\textrm{AB}}$ and $G_{\textrm{BA}}$ taking the second and third equations of the system \eqref{systemLangevin:appendix}. In the range $0\le t<d$, the differential equations for $G_{\textrm{AB}}(t)$ and $G_{\textrm{BA}}$ become homogeneous. The only solution for them satisfying the boundary conditions $G_{\textrm{AB}}(t=0)=0$ and $\dot{G}_{\textrm{AB}}(t=0)=0$ is $G_{\textrm{BA}}=G_{\textrm{AB}}=0$, which confirms that information cannot travel faster than light (recall that $c=1$ here). As a consequence, from the first and fourth of Eq.~\eqref{systemLangevin:appendix}, we can immediately see that the differential equations for the Green functions $G_{KK}(d\le t<2d)$ remain homogeneous also in the range $d\le t<2d$. Imposing the continuity of $G_{KK}$ and $\dot{G}_{KK}$ at $t=d$, we conclude that the solution for $G_{KK}$, given by Eq.~\eqref{GreenAA1st}, is valid also in the range $d\le t<2d$.

The second and third differential equation of the system \eqref{systemLangevin:appendix}, for $G_{\textrm{AB}}$ and $G_{\textrm{BA}}$ become non-homogeneous at $t>d$. From the time $d$ to $3d$, the non-homogeneous term is proportional to $G_{\textrm{BB}}(t-d)$ or $G_{\textrm{AA}}(t-d)$ given by Eq.~\eqref{GreenAA1st}. By imposing the continuity of $G_{\textrm{AB}}$ and $\Dot{G}_{\textrm{AB}}$, or $G_{\textrm{BA}}$ and $\Dot{G}_{\textrm{BA}}$, at $t=d$ we have that, in the range $d<t<3d$, the off-diagonal Green's function $G_{\textrm{AB}}=G_{\textrm{BA}}$ reads
    \begin{align}\label{greenABfirst:appendix}
        G_{\textrm{AB}}(t)=&\frac{2}{d}\frac{\sqrt{\gamma_\textrm{A}\gamma_\textrm{B}}\theta(t-d)}{\left(4\gamma_\textrm{A}\gamma_\textrm{B}(\Sigma_\textrm{A}^2+\Sigma_\textrm{B}^2)-4\Sigma_\textrm{B}^2\gamma_\textrm{A}-4\Sigma_\textrm{A}^2\gamma_\textrm{B}^2+(\Sigma_\textrm{A}^2-\Sigma_\textrm{B}^2)^2\right)}\nonumber\\
        &\times\left(\frac{\left(2\gamma_\textrm{A}^2+(\Sigma_\textrm{A}^2-\Sigma_\textrm{B}^2)-2\gamma_\textrm{A}\gamma_\textrm{B}\right)\sinh\left((t-d)\sqrt{\gamma_\textrm{A}^2+\Sigma_\textrm{A}^2}\right)}{\sqrt{\gamma_\textrm{A}^2+\Sigma_\textrm{A}^2}}e^{-\gamma_\textrm{A}(t-d)}+\textrm{A}\longleftrightarrow\textrm{B}\right.\nonumber\\
     &\left.+2(\gamma_\textrm{A}-\gamma_\textrm{B})\left(\cosh\left((t-d)\sqrt{\gamma_\textrm{A}^2+\Sigma_\textrm{A}^2}\right)e^{-\gamma_\textrm{A}(t-d)}-\cosh\left((t-d)\sqrt{\gamma_\textrm{B}^2+\Sigma_\textrm{B}^2}\right)e^{-\gamma_\textrm{B}(t-d)}\right)\right)\,.
    \end{align}
When the detectors are identical, i.e. $\gamma_\textrm{A}=\gamma_\textrm{B}$ and $\omega_\textrm{A}=\omega_B$, Eq.~\eqref{greenABfirst:appendix} reduces to
\begin{equation}\label{eq: transmission dec equals}
    G_{\textrm{AB}}(t)=\gamma\frac{ (t-d)\theta(t-d)e^{-\gamma (t-d)}}{d\gamma_\Sigma^2}
    \cosh(\gamma_\Sigma(t-d))
    \left(1-\frac{\tanh(\gamma_\Sigma(t-d))}{\gamma_\Sigma(t-d)}\right),
\end{equation}
where we have introduced $\gamma_\Sigma:\sqrt{\gamma^2+\Sigma^2}$ for convenience of presentation.

To this point we have solved the differential equations \eqref{systemLangevin:appendix} in the range of times $0\le t<2d$. The Green function solutions give the transmissivity and the noise in the same range of times. In principle, one can put the Green function \eqref{greenABfirst:appendix} into the first and fourth line of the system \eqref{systemLangevin:appendix} to calculate the Green functions $G_{KK}(t)$ at times $3d<t<5d$. Then, one can use the solutions obtained this way and insert them into the second and third of Eq.~\eqref{systemLangevin} computing $G_{\textrm{AB}}(t)=G_{\textrm{BA}}(t)$ at times $2d<t<4d$. One can continue with this procedure indefinitely, obtaining solutions for all the times. Analytical solutions are always expected for each range of times since the inhomogeneous terms appearing in the differential Eqs.~\eqref{systemLangevin:appendix} are always sums of exponentials. However, with increasing time $t$, these solutions become increasingly complicated. For all purposes, if we want to know the behaviour of the Green functions in an arbitrary range of time, numerical calculations are necessary and solving the system \eqref{systemLangevin:appendix} step by step is required.

We now show that, considering a particular range for the parameters $\omega_K$, $\gamma_K$, $d$ and $\sigma$, the equations for the Green functions \eqref{GreenAA1st} and \eqref{greenABfirst:appendix} that are valid at $0\le t<2d$ can be considered valid also at $t\ge 2d$. We do this by analyzing the equations in the system \eqref{systemLangevin:appendix}: if the right hand side of those equations is negligible, then the Green functions are approximately the solutions \eqref{GreenAA1st} and \eqref{greenABfirst:appendix} even for $t\ge2d$. This argument can be verified through the homogeneous Eq.~\eqref{HomLangevin} as follows. We can apply the Fourier transform on both sides and easily obtain 
\begin{equation}\label{FTgreen}
    \widetilde{G}_{KK}(z)=\frac{-\Sigma_K^2-z^2-2{\rm i}z\gamma_K}{\text{det}\,\widetilde{\mathbb{G}}(z)^{-1}}\,,
\end{equation}
and 
\begin{equation}
    \widetilde{G}_{ij}(z)=\frac{\frac{2\sqrt{\gamma_K\gamma_j}}{d}e^{{\rm i}z d}}{\text{det}\,\widetilde{\mathbb{G}}(z)^{-1}}\,,
\end{equation}
where we have defined $\text{det}\,\widetilde{\mathbb{G}}(z)^{-1}=
        \bigl(\Sigma_\textrm{A}^2+z^2+2{\rm i}z\gamma_\textrm{A}\bigr)\bigl(\Sigma_\textrm{B}^2+z^2+2{\rm i}z\gamma_\textrm{B}\bigr)-\frac{4\gamma_\textrm{A}\gamma_\textrm{B}}{d^2}e^{2{\rm i}dz }$.
If the condition $4\gamma_\textrm{A}\gamma_\textrm{B}\ll d^2|\Sigma_\textrm{A}^2\Sigma_\textrm{B}^2|$ holds, then the last term in the denominator can be neglected and the Fourier-transformed Green functions become
\begin{equation}\label{FTgreenAAsimplified}
    \widetilde{G}_{KK}(z)=-\frac{1}{\Sigma_K^2+z^2+2{\rm i}z\gamma_K},
\end{equation}
and 
\begin{equation}\label{FTgreenABsimplified}
    \widetilde{G}_{\textrm{AB}}(z)=-\frac{2\sqrt{\gamma_\textrm{A}\gamma_\textrm{B}}e^{{\rm i}dz}}{d\left(\Sigma_\textrm{A}^2+z^2+2{\rm i}z\gamma_\textrm{A}\right)\left(\Sigma_\textrm{B}^2+z^2+2{\rm i}z\gamma_\textrm{B}\right)},
\end{equation}
By computing the inverse Fourier transform of Eqs.~\eqref{FTgreenAAsimplified} and \eqref{FTgreenABsimplified} and imposing the causality condition, one obtains exactly the solutions \eqref{GreenAA1st} and \eqref{greenABfirst:appendix}, respectively. This proves that, when $4\gamma_\textrm{A}\gamma_\textrm{B}\ll d^2|\Sigma_\textrm{A}^2\Sigma_\textrm{B}^2|$, the solutions of Eq.~\eqref{systemLangevin:appendix} for the Green functions at times $t\ge 2d$ can be approximated to the ones at times $0\le t< 2d$, namely Eq.~\eqref{GreenAA1st} for $G_{KK}(t)$ and Eq.~\eqref{greenABfirst:appendix} for $G_{\textrm{BA}}(t)=G_{\textrm{AB}}(t)$.

The validity of this approximation in the range $|\Sigma_K^2|\gg2\sqrt{\gamma_\textrm{A}\gamma_\textrm{B}}/d$ can be seen by comparing Fig.~\ref{capacity1}, where the approximation is performed, with Fig.~\ref{capacity5} given by numerical calculations without performing the approximation. Indeed, from Fig.~\ref{capacity5}, we see that, by increasing $|\Sigma^2|$, the behaviour of the capacity converges to the one predicted with the approximation in Fig.~\ref{capacity1}. The  same behaviour can be seen by comparing Figs.~\ref{capacity3} and \ref{capacity6}.

\section{Noise calculation}\label{appendixB}
We can evaluate the noise produced by the channel through the determinant of the matrix $\mathbb{N}$ from 
\begin{align}\label{noisematrix:appendix}
    \mathbb{N}(t)=&T_{\textrm{BB}}(t)\sigma_{\textrm{BB}}(0)T_{\textrm{BB}}(t)^T+\int_0^tds\int_0^tds'\eta(t-s)\nu(s,s')\eta(t-s'),
\end{align}
which we denoted by $W\coloneqq\det\mathbb{N}$. To do that, we use the elements of the noise kernel \eqref{kerneldiag} and \eqref{convergentkernel}. Thus, starting from Eq.~\eqref{noisematrix:appendix}, we calculate the elements of $\mathbb{N}$.

The diagonal elements of the noise kernel \eqref{kerneldiag} leads to a divergence on the integral \eqref{noisematrix:appendix}. However, by using the equation \eqref{noisekernelRealD} for finite size detectors and computing numerically the integral in  Eq.~\eqref{noisematrix:appendix}, the result coincides to the one we analytically obtain by considering 
\begin{equation}\label{noise kernel approximated}
    \nu_{KK}(t-t')\sim-\frac{\delta(t-t')}{4\sqrt{2\pi}\sigma}\,.
\end{equation}
The reason is that, by applying the approximation \eqref{approximation}, we have
\begin{equation}
    -\frac{\delta(t-t')}{4\sqrt{2\pi}\sigma}=-\frac{1}{4\pi\sqrt{2\pi}\sigma}\frac{\sqrt{2\pi}\sigma}{\pi(t-t')^2+2\pi\sigma^2}\sim-\frac{1}{4\pi^2}\frac{1}{(t-t')^2}\,.
\end{equation}
In this way, Eq.~\eqref{noise kernel approximated} reduces exactly to Eq.~\eqref{kerneldiag} in the limit $\sigma\to0$. The divergence that Eq.~\eqref{kerneldiag} leads to the integral in Eq.~\eqref{noisematrix:appendix} is the same divergence it would occur by considering Eq.~\eqref{noise kernel approximated} in the limit $\sigma\to0$. We use then Eq.~\eqref{noise kernel approximated} for the diagonal elements of the noise kernel to obtain analytic solutions for the noise $W$.

The main contribution to the quantity $W$ comes from the terms including the diagonal elements of the noise kernel \eqref{kerneldiag}. The other terms are in fact smaller at least by a factor $\sigma/d\ll1$. The first term $T_{\textrm{BB}}\sigma_{\textrm{BB}}(0)T_{\textrm{BB}}$ of Eq.~\eqref{detN} has a comparable magnitude such that it can also be considered negligible when $\sigma\ll d$. At this point, an analytical solution is possible. In this appendix we report exclusively the one for identical detectors, which reads
    \begin{align}\label{detN}
          W=&\frac{1}{32 \pi  64 d^4 \gamma ^2 \gamma_\Sigma^{12} \Sigma ^8 \sigma ^2}\left[-4 \Sigma ^8 \left(e^{2 (d-t) \gamma } \gamma ^3 \theta (t-d) \left(\left(1+(t-d)^2 \gamma_\Sigma^2\right) \cosh \left(2 (t-d) \gamma_\Sigma\right)\right.\right.\right.\nonumber\\
          &\left.\left.-2 (d-t) \gamma_\Sigma \sinh \left(2 (d-t) \gamma_\Sigma\right)\right)+2 d^2 e^{-2 t \gamma } \gamma  \gamma_\Sigma^4 \sinh ^2\left(t \gamma_\Sigma\right)\right)^2\nonumber\\
          &+\left(\theta (t-d)\left(-e^{-2 (t-d) \gamma } \gamma ^3 \left(4 \gamma ^5-4 (t-d) \gamma ^4 \Sigma ^2-10 (t-d) \gamma ^2 \Sigma ^4-6 (t-d) \Sigma ^6+\gamma  \Sigma ^4 \left(9+2 (t-d)^2 \Sigma ^2\right)\right.\right.\right.\nonumber\\
          &\left.-\gamma ^3 \Sigma ^2 \left(-11-2 (t-d)^2\Sigma ^2\right)\right) \cosh \left(2 (t-d) \gamma_\Sigma\right)+e^{-2 (t-d) \gamma } \gamma ^3 \gamma_\Sigma \left(4 \gamma ^4-4 (t-d) \gamma ^3 \Sigma ^2\right.\nonumber\\
          &\left.\left.-8 (t-d) \gamma  \Sigma ^4+\Sigma ^4\left(5+2 (t-d)^2 \Sigma ^2\right)+\gamma ^2 \Sigma ^2 \left(9+2 (t-d)^2 \Sigma ^2\right)\right) \sinh \left(2 (d-t) \gamma_\Sigma\right)\right)\nonumber\\
          &\left.-2 d^2 e^{-2 t \gamma } \gamma\gamma_\Sigma^4 \Sigma ^4 \left(\gamma \cosh \left(2 t \gamma_\Sigma\right)+  \gamma_\Sigma \sinh \left(2 t \gamma_\Sigma\right)\right)\right)\nonumber\\
         & \times\left(\theta (-d+t) \left(-e^{2 (d-t) \gamma } \gamma^3 \left(-2 (-d+t) \gamma ^2 \Sigma ^2+2 (d-t) \Sigma ^4-\gamma  \Sigma ^2 \left(1+2 (t-d)^2 \Sigma ^2\right)+\gamma ^3 \left(1-2 (t-d)^2 \Sigma ^2\right)\right)\right.\right.\nonumber\\
          &\times\cosh \left(2 (d-t) \gamma_\Sigma\right)-e^{2 (d-t) \gamma } \gamma ^3 \gamma_\Sigma \left(-4 (-d+t) \gamma  \Sigma ^2-\Sigma ^2 \left(1+2 (t-d)^2 \Sigma ^2\right)\right.\nonumber\\
          &\left.\left.\left.\left.+\gamma ^2 \left(-1-2 (t-d)^2 \Sigma ^2\right) \sinh \left(2 (d-t) \gamma_\Sigma\right)\right)+2 d^2 \gamma  \gamma_\Sigma^4 \Sigma ^2 e^{-2 t \gamma }\left(\gamma  \cosh \left(2 t \gamma_\Sigma\right)-  \gamma_\Sigma \sinh
   \left(2 t \gamma_\Sigma\right)\right)\right)\right)\right]\,.
    \end{align}
The expression for the case of different detectors is even more complicated and we choose not to report it because it would not improve the understanding.

\section{Useful expressions}\label{appendixD}
In this appendix we report a few useful expressions in order to avoid encumbering the main text.

Through Eq. \eqref{classical capacity simplified} we showed that the late time capacity $C_{\infty}$, when $\Sigma^2>0$, is a monotonic function of the ratio $\frac{\tau}{\sqrt{W}}$. The latter is finite for $t\to\infty$ and is given by the following expression:
\begin{align}\label{Asymptotic:capacity:appendix}
\frac{\tau}{\sqrt{W}}= \sqrt{\frac{512\pi\sigma^2\Sigma ^8 \gamma^2\gamma_\Sigma}{\Sigma^4 \gamma^2\gamma_\Sigma+8 \gamma ^6 \gamma_\Sigma+12 \gamma ^5 \Sigma^2 +8 \gamma ^4 \Sigma^2  \gamma_\Sigma+8 \gamma ^7+4 \gamma^3  \Sigma^4+4 d^2
   \Sigma ^4 \gamma_\Sigma^3 \left(\gamma+\gamma_\Sigma\right)^2 e^{2 d \left(\gamma_\Sigma-\gamma \right)}}}.
\end{align}

From Eqs.~\eqref{GreenAA1st} and \eqref{greenABfirst:appendix}, we can obtain an analytic expression for transmissivity $\tau$ in the different detectors case, when $|\Sigma_i^2|\gg2\sqrt{\gamma_{\textrm{A}}\gamma_{\textrm{B}}}/d$, with $i=\{\textrm{A},\textrm{B}\}$. It reads
\begin{align}\label{eq: transm dec not equals}
    \tau=&\frac{4\gamma_\textrm{A}\gamma_\textrm{B}\theta(t-d)}{d^2\left(4\gamma_\textrm{A}\gamma_\textrm{B}(\Sigma_\textrm{A}^2+\Sigma_\textrm{B}^2)-4\Sigma_\textrm{B}^2\gamma_\textrm{A}^2-4\Sigma_\textrm{A}^2\gamma_\textrm{B}^2+(\Sigma_\textrm{A}^2-\Sigma_\textrm{B}^2)^2\right)}\nonumber\\&\times\left(e^{-2\gamma_\textrm{A}(t-d)}+e^{-2\gamma_\textrm{B}(t-d)}-e^{-(\gamma_\textrm{A}+\gamma_\textrm{B})(t-d)}\left(2\cosh\left(\gamma_{\Sigma_{\textrm{A}}}(t-d)\right)\cosh\left(\gamma_{\Sigma_{\textrm{B}}}(t-d)\right)\right.\right.\nonumber\\&\left.\left.-\frac{\Sigma_\textrm{A}^2+\Sigma_\textrm{B}^2+2\gamma_\textrm{A}\gamma_\textrm{B}}{\gamma_{\Sigma_{\textrm{A}}}\gamma_{\Sigma_{\textrm{B}}}}\sinh\left(\gamma_{\Sigma_{\textrm{A}}}(t-d)\right)\sinh\left(\gamma_{\Sigma_{\textrm{B}}}(t-d)\right)\right)\right)\,,
\end{align}
where we have defined $\gamma_{\Sigma_{\textrm{i}}}\coloneqq\sqrt{\gamma_i^2+\Sigma_i^2}$, with $i=\{\textrm{A},\textrm{B}\}$.

\end{document}